\documentclass{svproc}
\usepackage{url}

\usepackage{amsmath}
\usepackage{amssymb}
\usepackage{xcolor}
\usepackage[pdftex]{graphicx}
\usepackage{subcaption}
\usepackage{multirow}
\begin{document}
\mainmatter              % start of a contribution
\title{Particle filtering methods for partially observed branching processes}
\titlerunning{}  % abbreviated title (for running head)
\author{Miguel Gonz\'alez\inst{} \and In\'es M. del Puerto \and Manuel Serrano-Pastor}
\authorrunning{M. Gonz\'alez at al.} % abbreviated author list (for running head)
\tocauthor{Miguel Gonz\'alez,  In\'es M. del Puerto and Manuel Serrano-Pastor}
\institute{Department of Mathematics. Faculty of Sciences. Universidad de Extremadura, Badajoz 06006, SPAIN,\\
\email{mvelasco@unex.es, idelpuerto@unex.es, manuelsp@unex.es}}

\maketitle              % typeset the title of the contribution

\begin{abstract}
This paper focuses on the estimation of partially observed branching processes. First, the estimators from a frequentist perspective proposed in the literature are reviewed.  The main objective of this paper is to present  computational tools based on sequential Monte Carlo methods to perform Bayesian inference for these processes. In particular, the Liu-West particle filter is applied to perform Bayesian estimation of the parameters of interest for an epidemic model fitted by a partially observed branching process. As application, the example given in  \cite{MeesterUnknownTitle2006} is revisited and extended.
\end{abstract}
\section{Introduction}
The Bienaym\'e-Galton-Watson branching processes have been used to  approximate the beginning of an outbreak (see for instance \cite{ball95}, \cite{Becker1989} or \cite{Farrington2003}). In real-world epidemics, it is rare that every single infection is recorded. Some cases may be asymptomatic and go unnoticed, while others may not be reported due to lack of medical access, testing limitations, or misdiagnosis. What we experienced with COVID-19 pandemic is not far behind, during it both testing limitations and potential misdiagnoses played significant roles. Testing limitations included the availability of accurate tests, particularly in the early stages of the pandemic, and the potential for false negative results, especially with rapid antigen tests. Misdiagnosis, on the other hand, could stem from the similarity of COVID-19 symptoms to other illnesses, leading to delayed or inaccurate diagnoses (see for instance \cite{alvarez23}). Consequently, partially observed Bienaym\'e-Galton-Watson processes have been considered to deal with these situations. In this context a pioneer applied work with swine fever data  was done by \cite{Meester2002}, by considering particular finite supported offspring distributions. Later on this study was extended in \cite{MeesterUnknownTitle2006} under more general offspring distribution. Briefly, it is considered a population of infected individuals that are not observable and only a proportion of them are detected in a fixed interval of time, namely between times $n$ and $n+1$, independent of the others. Once an infectious individual is detected, it cannot continue spreading the disease. However, an infectious individual detected between times $n$ and $n + 1$ can transmit the disease from time $n$ until the moment of detection, acting independently of the others, being reasonable to assume a reduced number of offsprings respect to the one of an undetected individual. Both papers considered  individuals remain infectious until detected, as recovery or death cannot occur before detection. This assumption was dropped in \cite{Panaretos2007} and  \cite{KvitkoviovUnknownTitle2011}. Let us consider the model, $\{(X_n,Z_n))\}_{n\geq 0}$, introduced in \cite{Panaretos2007} given iteratively by
\begin{equation}\label{def}
X_0=1, \,\, Z_n=\sum_{i=1}^{X_n}\delta_{n i}, \,\, X_{n+1}=\sum_{i=1}^{X_n} X_{n i} (1-\delta_{n i}) +  \sum_{i=1}^{X_n} X'_{n i} \delta_{n i},\,\, n=0,1,\ldots,
\end{equation}
where all the variables are defined on the same probability space and $\{X_{n i}:n\geq 0, i\geq 1\}$ and $\{X'_{n i}:n\geq 0, i\geq 1\}$ are independent sequences 
of independent and identically distributed (i.i.d.) nonnegative integer-valued random variables; and $\{\delta_{ij}:i\geq 1,j\geq 0\}$ are a sequence of 
of i.i.d. Bernoulli random variables with success probability $\pi, 0<\pi<1$. All these variable are mutually independent. The model (\ref{def}) is referred as a partially observed branching process.

The partially observed branching process constitutes a versatile framework for modeling a variety of real-world phenomena. In the context of an epidemic, $X_n$ are the infectious individuals able to spread the disease and $Z_n$ are the detected individuals between times $n$ and $n+1$. Let denote $F_X$ the offspring cumulative probability distribution of a detected individual  (not being a degenerate distribution concentrated at 0) and 
$F_{X'}$ the one of the undetected individuals. Let us assume that both offspring distributions have finite fourth moments such that $F_X\leq F_{X'}$ (detected individuals may have a different offspring distribution than undetected ones, and assuming that they have less reproduction capacity than the undetected ones makes sense).
The probability of being detected is the same for all the individuals. The variables $Z_n$ are observed and $X_n$ are unobservable.

The process  $\{(X_n,Z_n)\}_{n\geq 0}$ is a Markov chain with stationary transition probabilities. Moreover it is not hard to check that 
$\{X_n\}_{n \geq 0}$ is a Bienaym\'e--Galton-–Watson process with offspring cumulative probability distribution the mixture of the distributions $F_X$ and $F_{X'}$ in proportion $1-\pi$ and $\pi$, respectively.

Let denote $m_+=E[X_{01}]$, $m_-=E[X'_{01}]$, $\sigma_+^2=Var[X_{01}]$ and   $\sigma_-^2=Var[X'_{01}]$. Consequently the offspring mean and variance  of the process $\{X_n\}_{n\geq 0}$, denoted by $m$ and $\sigma^2$, satisfy
\begin{eqnarray*}
m&=& \pi m_-+(1-\pi)m_+,	\\
\sigma^2&=&\pi\sigma^2_-+(1-\pi)\sigma^2_+
+\pi(1-\pi)(m_--m_+)^2.
\end{eqnarray*}
The key parameter of interest, as it significantly affects the behavior of a Bienaym\'e--Galton-–Watson branching process is the mean number of offspring, $m$. When the process is fully observed, the variance of the offspring distribution, $\sigma^2$, also plays an important role, especially in the estimation of the mean, since it relates to the variability of the estimators. Moreover, when modeling the progression of an epidemic using a partially observed branching process, accurately estimating $\pi$ is essential for formulating effective policies to stop the spread of the disease.

Meester and Trapman  \cite{MeesterUnknownTitle2006} examine the estimation of parameters associated with the moments of the offspring distribution under partial observation. They concentrated on the consistency of these estimators. It is not easy to derive the asymptotic distributions of their proposed estimators. Later on Kvitkovi\v{c}ov\'a and Panaretos \cite{KvitkoviovUnknownTitle2011}  propose alternative estimators and focus on their asymptotic distributions, with the aim of constructing asymptotic confidence intervals. They prove that they are consistent and asymptotically
Gaussian, all conditionally on the explosion set, namely $\{X_n\to\infty\}$.
These estimators are defined on the non-degenerate set, and their results rest on the assumption that the process never goes extinct—an assumption valid only for supercritical offspring means, i.e. $m>1$. On the other hand, Rahimov \cite{rahimov17a} considers the problem of estimation of parameters of the process with a large number of initial ancestors based on
partial observations. He addresses subcritical and critical cases, i.e. $m<1$ or $m=1$ for which the non-extinction assumption is not satisfied.

Problems related to statistical inference based on partially observed branching processes with stationary or non-stationary immigration were considered by Rahimov in \cite{rahimov17} and \cite{rahimov19}, respectively. Briefly, \cite{rahimov17} derives the conditions under which the estimator of the offspring mean is strongly consistent and asymptotically normal, based on partial observations of a subcritical process with i.i.d. immigration. However, in \cite{rahimov19}, results are obtained without imposing conditions on the criticality of the process, allowing for the development of a unified estimation procedure that does not require prior knowledge of the range of the offspring mean.

In \cite{xu19} statistical inference for partially observed branching processes with application to cell lineage tracking of in vivo hematopoiesis is considered. In particular, they
 estimate model parameters using
the generalized method of moments for partially observed continuous-time, multi-type branching processes.

In this paper, we address the Bayesian estimation of parameters in a partially observed branching process using particle filtering. To the best of our knowledge, this is the first study to apply a Bayesian framework for parameter estimation in such partially observed branching processes. Besides this Introduction, in Section \ref{S2} we present  an overview of estimators proposed in the literature for the parameters involved in partially observed branching processes without immigration from a frequentist outlook. Section \ref{s3} presents the Liu-West's algorithm. Finally, in Section \ref{S4} we deal with the problem of the estimation of the parameters in a partially branching process for a epidemic model.

\section{An overview of the proposed estimators from a frequentist viewpoint}\label{S2}

First, we will provide the estimators proposed in \cite{MeesterUnknownTitle2006}  for $m$, the first one based on two consecutive observable population sizes, $\{Z_{n-1}, Z_n\}$,  and the second on the sample based on the observable population sizes until generation $n$, $\{Z_0, \ldots, Z_n\}$:
\begin{equation}\label{estm}
\bar{m}_n = \displaystyle\frac{Z_{n}}{Z_{n-1}+1}, \quad 
\widetilde{m}_n = \displaystyle\frac{\sum_{i=1}^{n} Z_i}
{\sum_{i=0}^{n-1} Z_i}.
\end{equation}

\begin{theorem}\label{POBP_convergencia}
Let $\{(X_n,Z_n)\}_{n\geq 0}$ be the process defined in (\ref{def}). Then, as
    $n \to \infty$,
\begin{itemize}
    \item[(a)] $\bar{m}_n \to m$ a.s. on $\{X_n\to\infty\}$,
    \item[(b)] $\widetilde{m}_n \to m$ a.s. on $\{X_n\to\infty\}$.
    \end{itemize}
\end{theorem}
The proof of this result can be read in \cite{MeesterUnknownTitle2006}.
The estimators $\bar{m}_n$ and $\widetilde{m}_n$  are analogous to those studied  when the branching process is fully observed. As  their analogous in the full-observation model, these estimators also preserve the consistency. However, while the estimators in the fully--observed model are asymptotically Gaussian, the asymptotic distributions of the estimators in the partially-observed model are not easy to establish. This is because the dependence structure of the process $\{Z_k\}_{k \geq 0}$ is more complex than that of the process $\{X_k\}_{k \geq 0}$. In order to establish the asymptotic normality, in \cite{KvitkoviovUnknownTitle2011} it is proposed an alternative estimator based on $\{Z_0, \ldots, Z_n\}$. Let ${n}_0 = \lfloor (n - 1)/2 \rfloor$ and 
\begin{equation}
\widehat{m}_{n,o} = \displaystyle\frac{\sum_{k=0}^{{n}_0} Z_{2k+1}}{\sum_{k=0}^{{n}_0} Z_{2k}}.
\end{equation}

The following result establishes that the estimator $\widehat{m}_{n,o}$ is strongly consistent and retains the asymptotic properties of its analogue in the full observation case.

\begin{theorem}\label{ta}
Let $\{(X_n,Z_n)\}_{n\geq 0}$ be the model defined in (\ref{def}). Then, as 
    $n \to \infty$,
    \begin{itemize}
    \item[(a)]$\widehat{m}_{n,o} \to m$ 
    a.s. on $\{X_n\to\infty\}$,
    \item[(b)] 
$P \left( \gamma^{-1}
\sqrt{ \sum_{k=0}^{\widetilde{n}_0} Z_{2k}}(\widehat{m}_{n,o} - m) 
\leq x \middle| X_n\to\infty \right) \to \phi(x)$,  $x\in \mathbb{R}$,
\end{itemize}
with $\gamma^2 =  (1 - \pi)m +  \pi\sigma^2 + (1 + \pi) m^2 - 2\pi m m_-$ and $\phi(x)$ the cumulative distribution function of a standard normal distribution.
\end{theorem}
 The proof of this result can be read in \cite{KvitkoviovUnknownTitle2011}.

Notice that the asymptotic variance of the mean estimator (see (b) in Theorem \ref{ta})  depends on $\gamma$ which is a parameter
   that involves the parameter $\pi$. 
 So that if an estimator is proposed for $\gamma^2$, with the ones proposed for $m$ and further assumptions between the relation between $m$ and $m_-$, the parameter $\pi$ could be also estimate (see more details in Section \ref{S4}).

The proposed estimator for $\gamma^2$ in \cite{MeesterUnknownTitle2006} is
\begin{equation}\label{estgamma}\widehat{\gamma}_n=\frac{1}{n}\sum_{k=1}^n (Z_{k-1}+1)(\bar{m}_k-\widehat{m}_n)^2,\end{equation} also similar to those for the offspring varaince of a fully--observed branching process.
In \cite{MeesterUnknownTitle2006} the weak consistency of this estimator was established. In order to establish the asymptotic normality it is  
proposed in \cite{KvitkoviovUnknownTitle2011} as an also weak consistent estimator for $\gamma^2$,
\begin{equation}
\widehat{\gamma}^2_{n,o} = \frac{1}{{n}_0} \sum_{k=0}^{{n}_0} (Z_{2k} + 1) (\widetilde{m}_{2k+1} - \widehat{m}_{n,o})^2,
\end{equation}
that verifies, as $n\to\infty$,
\begin{equation}
P \left( (2\gamma^{4})^{-1/2}
\sqrt{{n}_0}(\widehat{\gamma}^2_{n,o} - \gamma^2) 
\leq x \middle| X_n\to\infty \right) \to \phi(x), \ x\in \mathbb{R}.
\end{equation}

\section{Bayesian Particle Filter: Liu-West algorithm}\label{s3}
In this section we consider the  sequential estimation of the states (filtering) and  the joint estimation of states and fixed parameters in a general dynamic state-space model. This provides the necessary context for the development of the Bayesian inference of the parameters of interest in partially observed branching processes in Section \ref{S4}. In particular, we describe the Liu--West filter.

Consider a Markovian dynamic model in which we observe a sequence of observation vectors $y_t$, for $t = 1, 2, \ldots$. Let $x_t$ denote the latent state vector at time $t$, and let $\theta$ be a fixed parameter of the model. The model is fully specified at each time step by two components:

\begin{itemize}
  \item An observation model defined by the conditional probability distribution:
  \begin{equation}
  p(y_t \mid x_t, \theta),
  \end{equation}
  
  \item A state transition model (or evolution equation), governed by the Markovian transition probability:
  \begin{equation}
  p(x_t \mid x_{t-1}, \theta).
  \end{equation}
\end{itemize}

Under this framework, each observation $y_t$ is conditionally independent of past states and observations given the current state $x_t$ and parameter $\theta$. Similarly, the latent state $x_t$ is conditionally independent of all past states and observations given $x_{t-1}$ and $\theta$. The partially observed branching process given by (\ref{def}) fits naturally with these assumptions.

Given an initial state distribution $p(x_0 \mid \theta)$ and a prior distribution $p(\theta)$ over the parameter space, the objective is to infer the joint posterior distribution:
\[
p(x_t, \theta \mid y^t),
\]
where $y^t = (y_1, \ldots, y_t)$ denotes the set of observations up to time $t$.

To approximate this posterior distribution in a sequential manner, we employ a Sequential Monte Carlo (SMC) method, which iteratively updates a weighted Monte Carlo sample representing $p(x_t, \theta \mid y^t)$. At each time step $t$, the posterior is approximated by a set of particles (discrete sample) and associated weights (not necessarily uniform). Upon receiving a new observation $y_{t+1}$, the algorithm updates this approximation to obtain a sample from the posterior distribution at the next time point, $p(x_{t+1}, \theta \mid y^{t+1})$.

Specifically we use the Liu and West's filter, a very general filter that can be considered a benchmark (see \cite{Liu-West2001} and \cite{Carvalho2010}).\\

{\bf Liu--West's algorithm:}

% Suppose we have a Monte Carlo sample $(x_t^{(i)},\theta_t^{(i)})$ and weights $\omega_t^{(i)}$, $i=1\ldots,N$, representing the joint posterior distribution $p(x_t,\theta \mid y^t)$. Here the suffix $t$ on the parameter sample indicates time $t$ posterior, not time-variation. If we introduce the information of $y_{t+1}$,  we are interested in to obtain a sample of size $N$ from $p(x_{t+1},\theta \mid y^{t+1})$.

Assume we have a Monte Carlo sample $\{(x_t^{(i)}, \theta_t^{(i)})\}_{i=1}^N$ with associated \linebreak weights $\{\omega_t^{(i)}\}_{i=1}^N$ representing the joint posterior distribution $p(x_t, \theta \mid y^t)$. Note that the subscript $t$ on the parameter $\theta_t^{(i)}$ denotes that the sample corresponds to the posterior at time $t$, rather than implying time dependence of the parameter itself. Upon incorporating the new observation $y_{t+1}$, our goal is to generate a sample of size $N$ approximating the updated posterior distribution $p(x_{t+1}, \theta \mid y^{t+1})$.

According with Bayes'rule
\begin{equation}\label{param_update}
\begin{split}
p(x_{t+1},\theta|y^{t+1})
&\propto p(y_{t+1}|x_{t+1},\theta)p(x_{t+1},\theta|y^{t})\\
&\propto p(y_{t+1}|x_{t+1},\theta)p(x_{t+1}|\theta,y^{t})p(\theta|y^{t}).
\end{split}
\end{equation}
% Using Monte Carlo approximation 
% \begin{equation}\label{state_update1}
% \begin{split}
% p(x_{t+1},\theta|y^{t+1})\propto
% \sum_{i=1}^{N} \omega_t^{(i)} p(y_{t+1}|x_{t+1},\theta_t^{(i)})p(x_{t+1}|x_t^{(i)}, \theta_t^{(i)})p(\theta|y^{t}).
% \end{split}
% \end{equation}

% For each $i=1,\ldots,N$, choose an estimate
% $\mu_{t+1}^{(k)}$ de $x_{t+1}$, as for instance the expected mean of the density
% $p(x_{t+1}|x_t^{(i)}, \theta_t^{(i)})$. :
% \begin{equation*}
% \begin{split}
% g_{t+1}^{(k)}\propto  
% \omega_t^{(k)} p(y_{t+1}|\mu_{t+1}^{(k)}).
% \end{split}
% \end{equation*}

% Un valor grande de $g_{t+1}^{(k)}$ indica que la partícula 
% $x_t^{(k)}$, cuando ``evoluciona'' del tiempo $t$ al tiempo $t+1$,
% es más verosímil que sea consistente con el dato $y_{t+1}$ que
% no lo sea. 

% Luego, los indices $j$ se muestrean con probabilidades proporcionales a $g_{t+1}^{(j)}$ y se extraen las partículas $x_{t+1}^{(j)}$ del estado actual de $p(x_{t+1}|x_t^{(j)})$ basándose en estos índices ``auxiliares". Estas partículas de estado muestreadas son esencialmente muestras por importancia de la densidad a posteriori en el tiempo $t+1$ y tienen los pesos asociados:
% \begin{equation*}
% \begin{split}
% \omega_{t+1}^{(j)}=  
% \frac{p(y_{t+1}|x_{t+1}^{(j)})}{p(y_{t+1}|\mu_{t+1}^{(j)})}.
% \end{split}
% \end{equation*}

Let $\overline{\theta}_t$ and $V_t$ be the posterior Monte Carlo mean and covariance matrix of the density $p(\theta|y^{t})$ respectively, calculated from Monte Carlo sample $\{(x_t^{(i)},\theta_t^{(i)})\}_{i=1}^N$ with weights $\{\omega_t^{(i)}\}_{i=1}^N$, namely,
\begin{equation*}
\begin{split}
\overline{\theta}_t &= \sum_{i=1}^N \theta_t^{(i)} \omega^{(i)}_t,
\\
V_t&=\sum_{i=1}^N\omega^{(i)}_t(\theta_t^{(i)}-\overline{\theta}_t)
(\theta_t^{(i)}-\overline{\theta}_t)^T.
\end{split}
\end{equation*}

A kernel smoothing strategy to approximate $p(\theta|y^{t})$ is given by 

\begin{equation}\label{nuc}
\begin{split}
p(\theta|y^{t})\approx
\sum_{i=1}^N\omega^{(i)}_tN(\theta|m_t^{(i)},h^2V_t),
\end{split}
\end{equation}
where
$N(\cdot|m,S)$  is the density of a multivariate normal with mean $m$ and variance–covariance matrix $S$. Consequently, in (\ref{nuc}) a mixture of independent normal distributions, $N(\theta|m_t^{(i)},h^2V_t)$, weighted by the sample weights $\omega^{(i)}_t$ is considered.  The kernel locations $m_t^{(i)}$ are given by
$$m_t^{(i)}=a \theta_t^{(i)}+ (1-a)\bar \theta_t,$$ with $a=(3\delta-1)/2\delta$ where $0< \delta \leq 1$ (usually $\delta=0.95$). The constant $a$ measures the extent of shrinkage. Finally $h^2=1-a^2$, with the constant $h$ a measurement of the degree of overdispersion (see \cite{Liu-West2001} for a detailed discussion).

Finally, dealing again with (\ref{param_update}) and considering the approach (\ref{nuc}), an extended version of the auxiliary particle filter algorithm, incorporating the parameter with the state (see \cite{Liu-West2001} for details) is considered.\\

As is described in \cite{Liu-West2001}, the algorithm has the following steps:

\begin{enumerate}
\item[]{\bf Step 1.} For each $i=1,\ldots, N$, identify the prior point estimates of $(x_t, \theta)$ given by $\left(\mu_{t+1}^{(i)},m_{t}^{(i)}\right)$ with
$$\mu_{t+1}^{(i)}=E\left[x_{t+1} \mid x_t^{(i)},\theta_t^{(i)}\right]$$ computed from the state transition distribution.
\\

\item[]{\bf Step 2.} Sample an auxiliary integer variable from the set $\{1,\ldots,N\}$ with probabilities proportional to 
$$\widetilde{\omega}_{t+1}^{(i)}\propto \omega_t^{(i)} p(y_{t+1}\mid \mu_{t+1}^{(i)}, m_t^{(i)});$$ call the sampled index $k$.\\

\item[]{\bf Step 3.} Sample a new parameter vector $\theta_{t+1}^{(k)}$ from the $k^{th}$ normal component of the kernel density, namely $$\theta_{t+1}^{(k)} \sim N(\cdot \mid m_t^{(k)},h^2 V_t)$$ \\

\item[]{\bf Step 4.} Sample a value of the current state vector $x_{t+1}^{(k)}$ from the evolution equation $$x_{t+1}^{(k)}\sim p(\cdot \mid x_{t}^{(k)}, \theta_{t+1}^{(k)}).$$\\

\item[]{\bf Step 5.} Evaluate the corresponding weight $${\omega}_{t+1}^{(k)}\propto \frac{p(y_{t+1}\mid x_{t+1}^{(k)}, \theta_{t+1}^{(k)})}{p(y_{t+1}\mid \mu_{t+1}^{(i)}, m_t^{(i)})}.$$\\

\item[]{\bf Step 6.} Repeat steps 2-5 $N$ times to produce a sample $\{(x_{t+1}^{(i)}, \theta_{t+1}^{(i)})\}_{i=1}^N$ with associated weights $\{\omega_{t+1}^{(i)}\}_{i=1}^N$ representing the joint posterior distribution $p(x_{t+1}, \theta \mid y^{t+1})$.\\

\end{enumerate}

\section{Application to an epidemic model}\label{S4}

To illustrate the methodology we analyse from the Bayesian outlook the example given in \cite{MeesterUnknownTitle2006}. The example considers the discrete approximation of the standard SIR (susceptible-infective-removed) epidemic, where the population of susceptibles is considered to be infinite. The evolution of the number of infected individuals is modelled through a partially observed branching process. An undetected infective individual can cause (in the time interval that define the generations, e.g. weeks) new infections according to a Poisson distribution of parameter $\lambda$. An infective individual that is not detected remains infective itself. This implies that $X_{ni}$ is equal in distribution to $1 + X''_{ni}$ with $X''_{ni}$ following a Poisson distribution with parameter $\lambda$ and verifying the same conditions of independence and identical distribution as the variables $X_{ni}$. Hence, $m_+ =\lambda+1$ and $\sigma^2_+ = \lambda$. Respect to modelling the behaviour of the detected individuals in the corresponding interval, its is assumed that a detected individual is infective during a (known) fraction of the detection interval, $\phi$. Then $m_-=\phi \lambda$ and $\sigma_-^2=\phi \lambda$. Hence,
 \begin{align}
  m & =  (1-\pi) (\lambda +1)+\phi \lambda \pi, \label{eq1}\\
\sigma^2 & =  (1-\pi) \lambda+\phi \lambda \pi + \pi(1-\pi) ((1-\phi)\lambda + 1)^2,\nonumber\\
\gamma^2 &= (1-\pi)\lambda+(1-\pi)^2+(1-\pi)(\lambda +1)^2+\phi \lambda \pi, \label{eq2}
 \end{align}
where, recall, $\pi$ is the probability that an infective individual is detected. Assuming that the detection time is uniformly distributed over the interval of detection, then $m_- = \lambda/2$ seems reasonable and, then, we can fix $\phi=1/2$.

In \cite{Meester2002} an analysis of the data from the 1997 Dutch classical swine fever outbreak is developed. This study is retrieved and revisited in \cite{MeesterUnknownTitle2006}. Specifically, with the data of the Stage 2 of the outbreak, they develop a simulation study using the maximum likelihood estimates (MLEs) obtained in \cite{Meester2002} as the real parameter values in order to analyse the behaviour of the estimators of $m$ and $\pi$ that they propose (see Section \ref{S2}, notice here that given the relationship between $m$ and $m_{-}$, the estimation of $\pi$ can be obtained by using equations (\ref{eq1}) and (\ref{eq2}), and estimates given by (\ref{estm}) and (\ref{estgamma})). The authors are particularly interested in study the speed of convergence of the estimator of $\pi$, indicating that even after 750 generations (weeks), $\pi$ was not accurately estimated. They use the values $\pi=0.4$ and $\lambda=0.6$ (these are the MLEs in \cite{Meester2002}), and, hence $m=1.08$. 

We are interested in applying Bayesian inference based on particle filtering methods in this setting. We have extended the simulation study provided in \cite{MeesterUnknownTitle2006} by considering that $\pi$ takes more values between 0 and 1 not only 0.4. Moreover we have considered that the growth rate of the process is constant equal to $m=1.08$ (so that the results are comparable). Specifically, we have considered partially observed branching processes with $x_0=100$, $m=1.08$, $\phi=0.5$, $\pi \in \{0.2,0.4,0.6,0.8\}$ and with $\lambda$ determined by the relation 
\begin{equation} \label{eq.lambda}
\lambda=\frac{m-(1-\pi)}{1-\pi + \phi \pi},
\end{equation}
hence being $\lambda=0.31, 0.6, 0.97, 1.47$, respectively. Notice that the case studied in \cite{MeesterUnknownTitle2006} with $(\pi,\lambda)=(0.4,0.6)$ is also considered. For each one of these scenarios we have simulated a sample of size $n=30$ (see Table \ref{t1} for the specific samples). To make the simulations it is convenient to rewrite the model (\ref{def}) as:
\begin{align*}
&X_0=x_0\\
&Z_n \sim {Binomial}(X_n,\pi),\\ &X_{n+1}=\sum_{i=1}^{X_n-Z_n} X_{n i}  +  \sum_{i=1}^{Z_n} X'_{n i},\,\, n=0,1,\ldots
\end{align*}
Thus, $X_{n+1}| (X_n=x,Z_n=z, \theta)$ is equal in distribution to $x-z + Y$ with $Y$ following a Poisson distribution with parameter $\lambda(x-z+\phi z)$.
{
\begin{table}[h!]
\centering
\footnotesize
\renewcommand{\arraystretch}{1.5}
\setlength{\tabcolsep}{7pt}
\begin{tabular}{|c||c|c||c|c||c|c||c|c||c|c|}
\hline
& \multicolumn{2}{c||}{$\pi=0.2$}
& \multicolumn{2}{c||}{$\pi=0.4$}
& \multicolumn{2}{c||}{$\pi=0.6$}
& \multicolumn{2}{c||}{$\pi=0.8$}
& \multicolumn{2}{c|}{$\pi=0.4$}
\\ 
& \multicolumn{2}{c||}{$\lambda=0.31$}
& \multicolumn{2}{c||}{$\lambda=0.6$}
& \multicolumn{2}{c||}{$\lambda=0.97$}
& \multicolumn{2}{c||}{$\lambda=1.47$}
& \multicolumn{2}{c|}{$\lambda=0.6$}
\\ \hline
\textbf{$n$} &
\textbf{\( X_n \)} & \textbf{\( Z_n \)} & \textbf{\( X_n \)} & \textbf{\( Z_n \)} & \textbf{\( X_n \)} & \textbf{\( Z_n\)} & \textbf{\( X_n \)} & \textbf{\( Z_n \)} & \textbf{\( X_n \)} & \textbf{\( Z_n \)} \\ \hline
0 &100 & 13 & 100 & 44 & 100 & 60 & 100 & 81 & 100 & 41 \\ \hline
1&111 & 25 & 105 & 34 & 103 & 61 & 102 & 88 & 99  & 42 \\ \hline
2&111 & 28 & 122 & 46 & 113 & 78 & 99  & 79 & 109 & 42 \\ \hline
3&109 & 20 & 148 & 56 & 98  & 60 & 99  & 83 & 125 & 53 \\ \hline
4&124 & 37 & 162 & 59 & 111 & 69 & 100 & 80 & 134 & 52 \\ \hline
5&111 & 23 & 172 & 79 & 113 & 72 & 109 & 90 & 150 & 54 \\ \hline
6&127 & 23 & 172 & 57 & 126 & 75 & 120 & 96 & 157 & 64 \\ \hline
7&142 & 28 & 180 & 62 & 141 & 81 & 125 & 96 & 187 & 73 \\ \hline
8&157 & 36 & 205 & 80 & 150 & 92 & 153 & 122 & 204 & 85 \\ \hline
9&159 & 26 & 208 & 88 & 163 & 104 & 173 & 132 & 218 & 80 \\ \hline
10&179 & 30 & 222 & 76 & 180 & 108 & 190 & 149 & 241 & 103 \\ \hline
11&193 & 44 & 273 & 100 & 190 & 111 & 216 & 173 & 238 & 90 \\ \hline
12&193 & 42 & 304 & 139 & 217 & 132 & 217 & 179 & 258 & 100 \\ \hline
13&209 & 47 & 308 & 117 & 232 & 149 & 238 & 189 & 288 & 127 \\ \hline
14&206 & 42 & 333 & 148 & 247 & 151 & 257 & 206 & 294 & 114 \\ \hline
15&218 & 35 & 353 & 144 & 263 & 162 & 275 & 223 & 321 & 135 \\ \hline
16&263 & 47 & 392 & 155 & 264 & 157 & 279 & 231 & 333 & 141 \\ \hline
17&287 & 55 & 415 & 161 & 277 & 165 & 289 & 237 & 357 & 120 \\ \hline
18&323 & 70 & 451 & 187 & 325 & 181 & 337 & 269 & 410 & 154 \\ \hline
19&348 & 61 & 450 & 199 & 371 & 213 & 360 & 297 & 433 & 176 \\ \hline
20&392 & 80 & 452 & 178 & 423 & 266 & 353 & 287 & 483 & 193 \\ \hline
21&412 & 87 & 470 & 185 & 413 & 233 & 389 & 312 & 518 & 211 \\ \hline
22&443 & 105 & 495 & 198 & 437 & 256 & 411 & 331 & 555 & 223 \\ \hline
23&457 & 95 & 546 & 204 & 479 & 296 & 431 & 348 & 607 & 264 \\ \hline
24&500 & 102 & 593 & 228 & 491 & 302 & 445 & 372 & 613 & 250 \\ \hline
25&550 & 98 & 663 & 270 & 513 & 307 & 463 & 383 & 659 & 268 \\ \hline
26&599 & 121 & 713 & 304 & 570 & 348 & 500 & 406 & 714 & 291 \\ \hline
27&633 & 142 & 742 & 311 & 615 & 389 & 533 & 428 & 783 & 302 \\ \hline
28&686 & 136 & 807 & 352 & 610 & 371 & 602 & 482 & 886 & 327 \\ \hline
29&744 & 139 & 854 & 349 & 628 & 379 & 654 & 515 & 984 & 411 \\ \hline
30&817 & 173 & 922 & 360 & 684 & 392 & 722 & 573 & 1030 & 416 \\ \hline
\end{tabular}
\vspace*{0.25cm}

\caption{Simulated samples.}\label{t1}
\end{table}
}

To apply the Liu-West algorithm we consider the parameter vector $\theta=(\pi,\lambda) \in (0,1)\times (0,\infty)$, the observed sample given by $\{Z_0,\ldots Z_n\}$ and need to know the distribution of $Z_{t+1}|(X_{t+1}, \theta)$ which is binomial of size $X_{t+1}$ and success probability equal to $\pi$, and the transition distribution $X_{t+1}|(X_t,\theta)$ of the Markov chain $\{X_n\}_{n \geq 0}$. To obtain this last we take advantage of the fact that $\{X_n\}_{n \geq 0}$ is a Bienaym\'e-Galton-Watson process and use Monte Carlo methods to sample from the transition distributions.

For all the examples we have considered sets of $N=2000$ particles. Initially we assume that $x_0=100$ is known. As prior for $\theta$ we consider an uniform distribution on the curve  $$\lambda=\frac{\widetilde{m}_n-(1-\pi)}{1-\pi + \phi \pi}.$$ To do that it is only needed to sample $\pi$ from an uniform distribution in the interval $(0,1)$ and apply last equation to obtain the corresponding values of $\lambda$. We take advantage of the fact that $\widetilde{m}_n$ is a strongly consistent estimator on $m$. This prior shows that we have not relevant information on the possible values of $\pi$ and $\lambda$, beyond its relation given by equation (\ref{eq.lambda}).

\begin{figure}[ht]
    \centering
    % Primera fila de imágenes
    \begin{subfigure}[b]{0.45\textwidth}
        \includegraphics[width=\textwidth]
{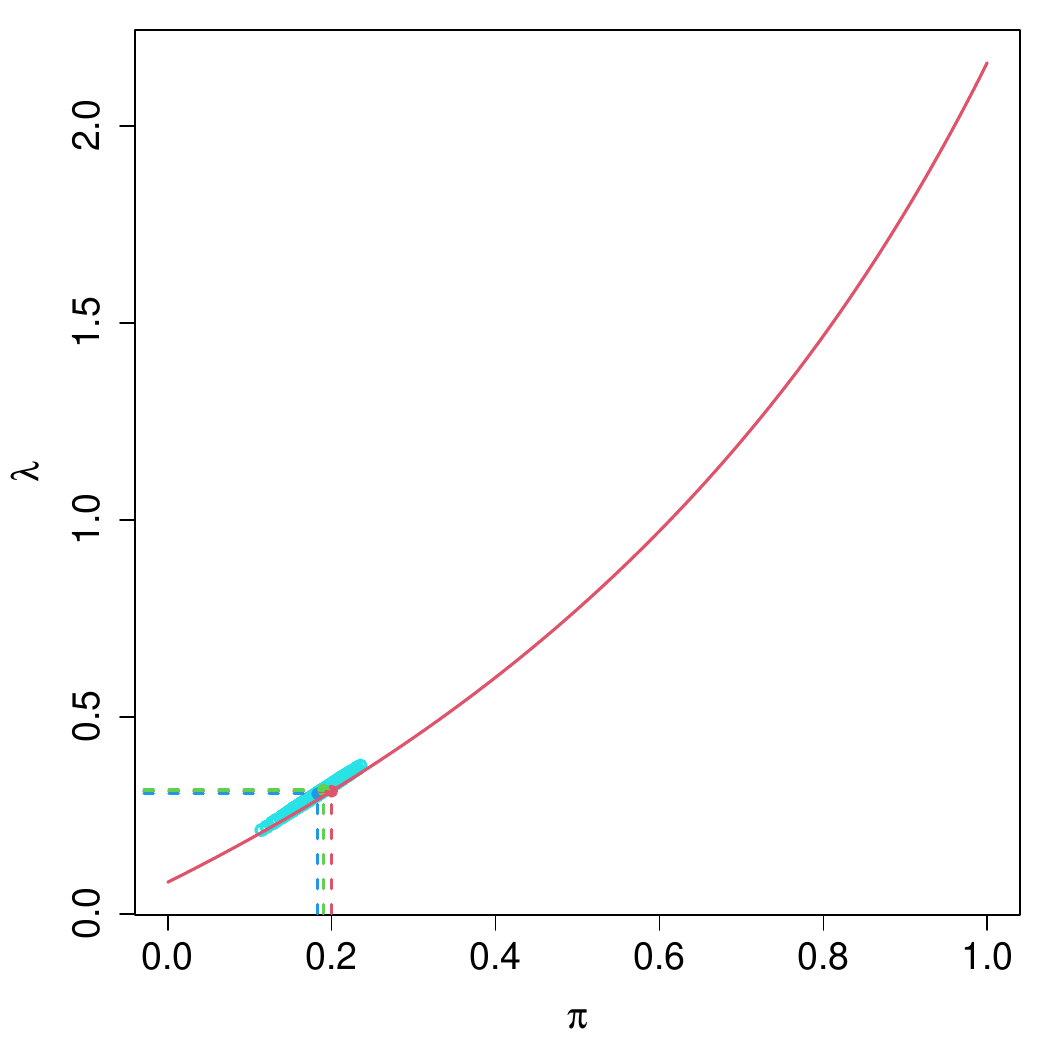}
       \caption{Scatterplot of the particles sampled from the joint posterior density estimate of $\theta=(\pi,\lambda$), its posterior mean (blue dot), posterior mode (green dot) and the true parameter value (red dot). The red curve is given by equation (\ref{eq.lambda}).}
        \label{fig:subfig1}
    \end{subfigure}
    \hfill
    \begin{subfigure}[b]{0.45\textwidth}
        \includegraphics[width=\textwidth]
{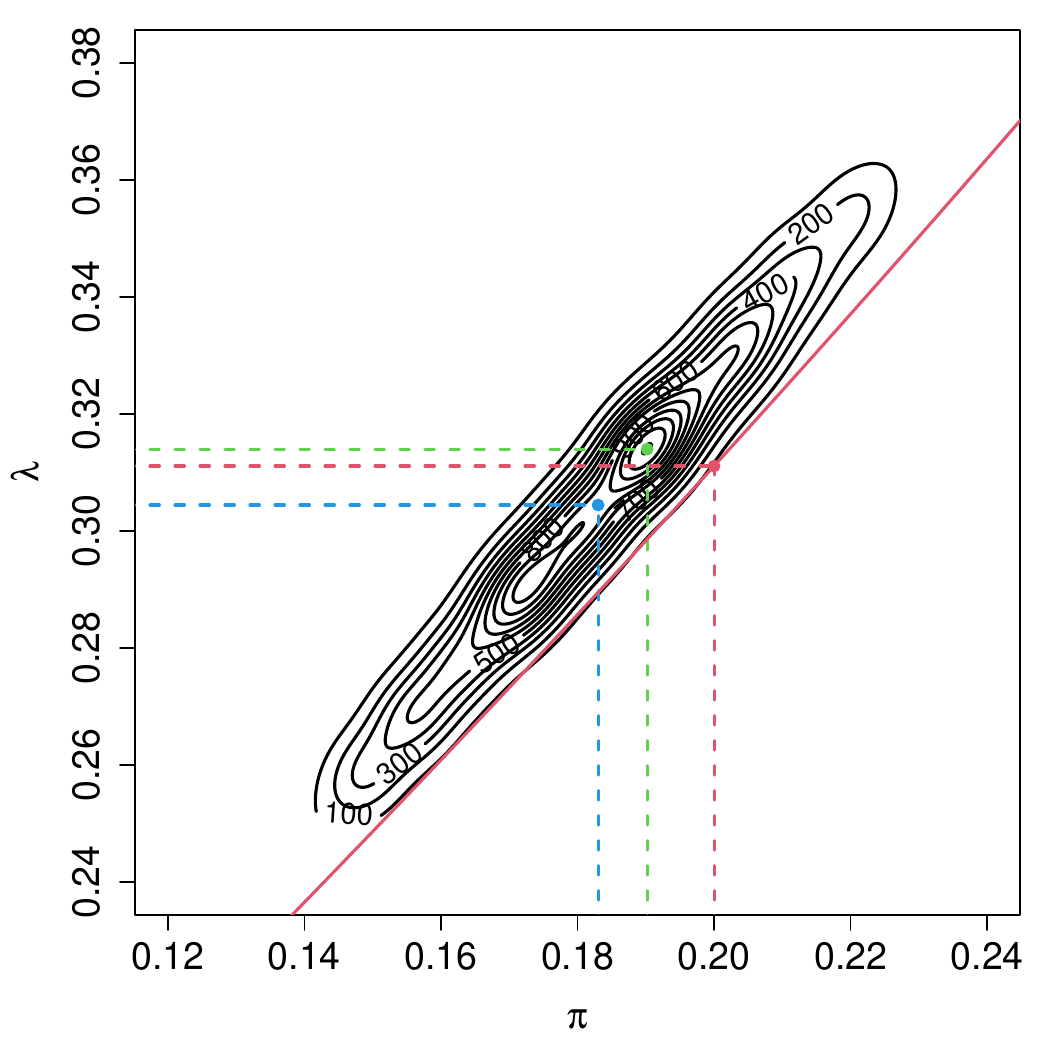}
        \caption{Contour plot of the joint posterior density estimate of $\theta=(\pi,\lambda$), its posterior mean (blue dot), posterior mode (green dot) and the true parameter value (red dot). The red curve is given by equation (\ref{eq.lambda}).}
        \label{fig:subfig2}
    \end{subfigure}

    \vspace{1cm}

\begin{subfigure}[b]{0.45\textwidth}
        \includegraphics[width=\textwidth]
{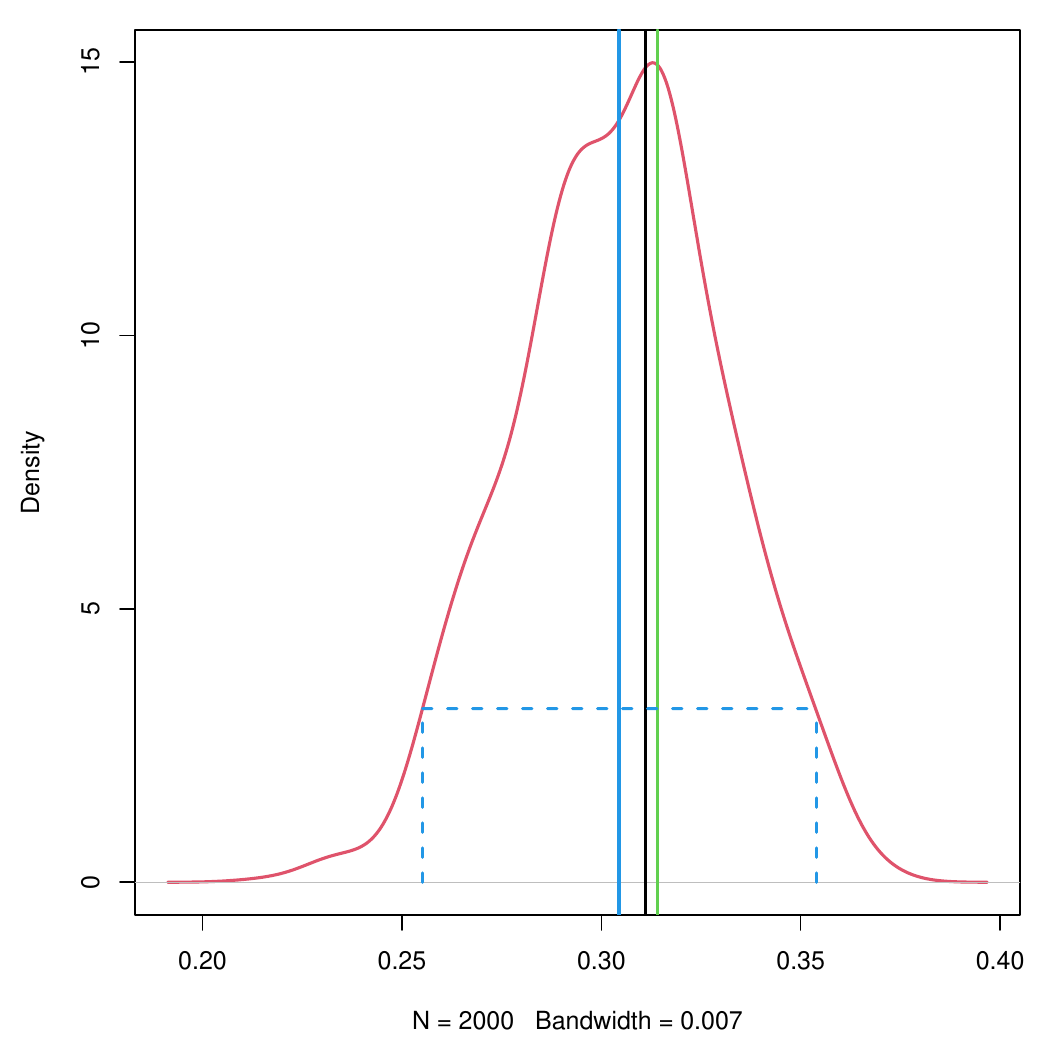}
        \caption{Marginal posterior density estimate of $\lambda$ with its posterior mean (blue line), posterior mode (green line), the true parameter value (black line) and the 95\% HPD credible set.}
        \label{fig:subfig3}
    \end{subfigure}
    \hfill
    \begin{subfigure}[b]{0.45\textwidth}
        \includegraphics[width=\textwidth]
{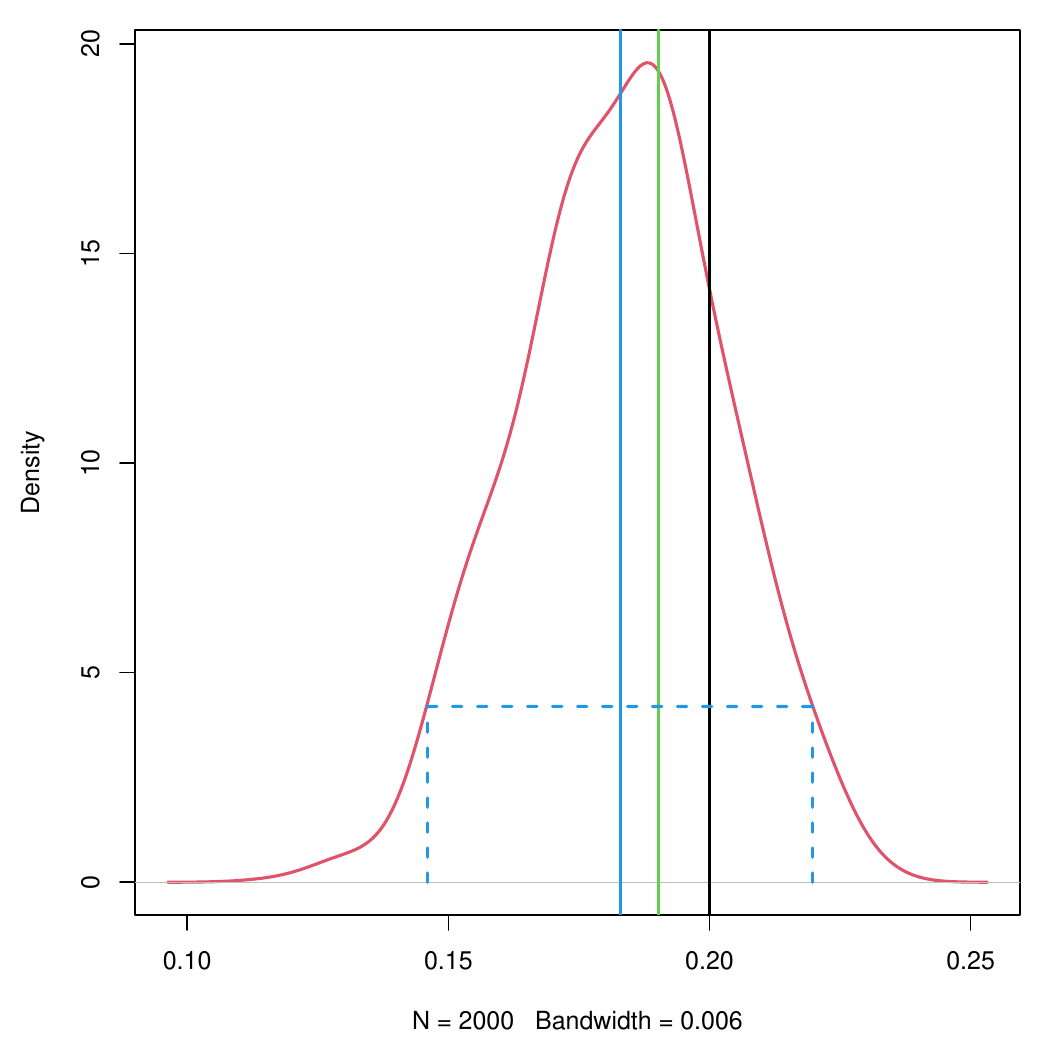}
        \caption{Marginal posterior density estimate of $\pi$ with its posterior mean (blue line), posterior mode (green line), the true parameter value (black line) and the 95\% HPD credible set.}
        \label{fig:subfig4}
    \end{subfigure}

    \caption{Case: $\pi=0.2$ and $\lambda=0.31$.}
    \label{fig:main1}
\end{figure}

\begin{figure}[ht]
    \centering
    % Primera fila de imágenes
    \begin{subfigure}[b]{0.45\textwidth}
        \includegraphics[width=\textwidth]
{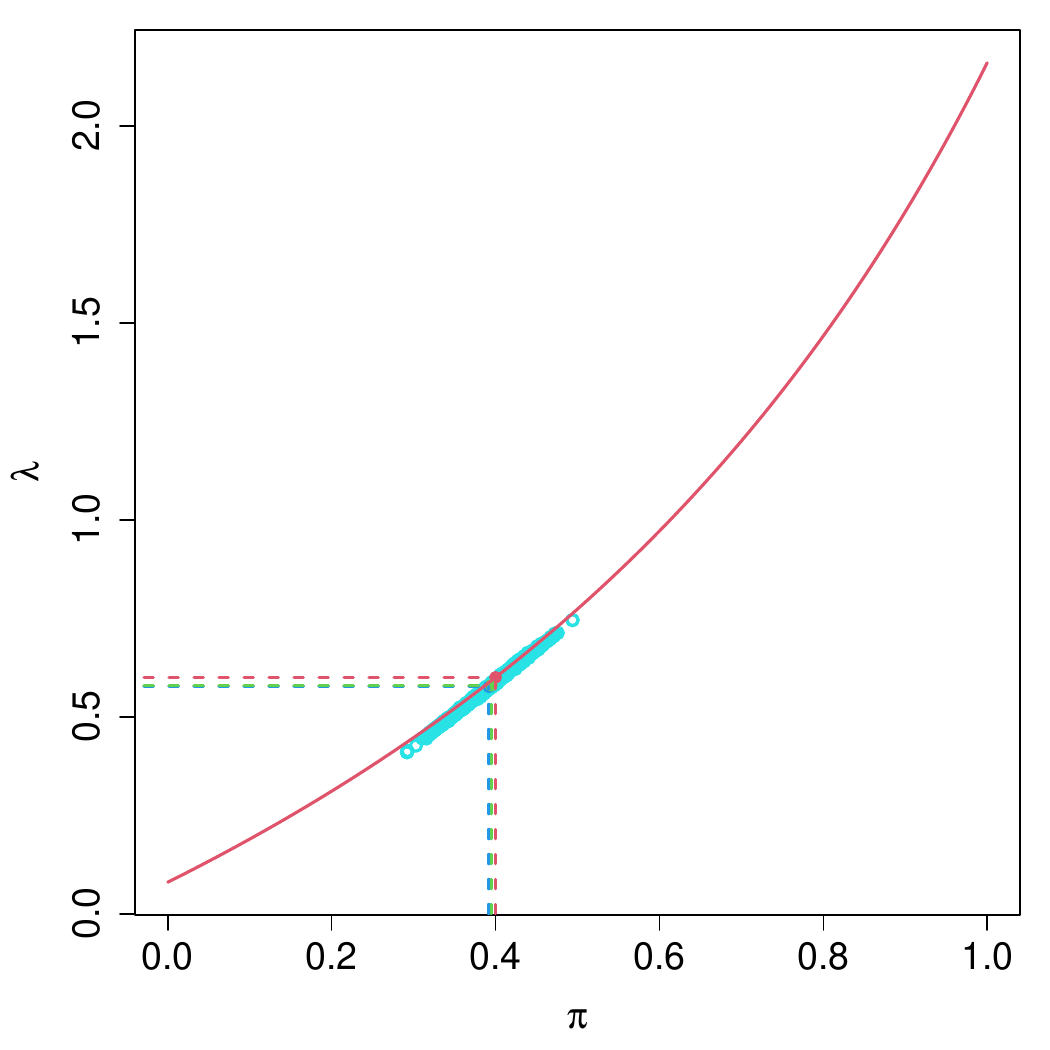}
        \caption{Scatterplot of the particles sampled from the joint posterior density estimate of $\theta=(\pi,\lambda$), its posterior mean (blue dot), posterior mode (green dot) and the true parameter value (red dot). The red curve is given by equation (\ref{eq.lambda}).}
        \label{fig:subfig5}
    \end{subfigure}
    \hfill
    \begin{subfigure}[b]{0.45\textwidth}
        \includegraphics[width=\textwidth]
{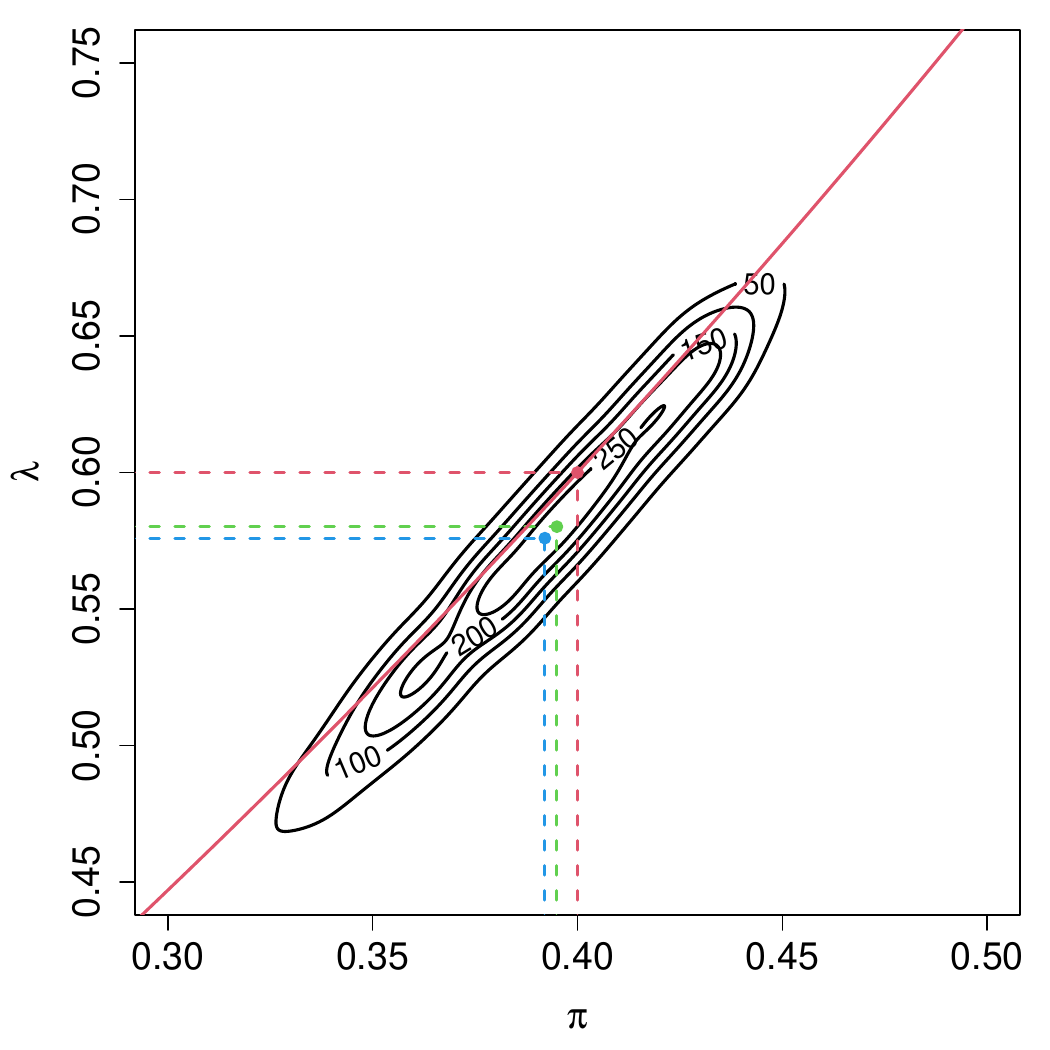}
        \caption{Contour plot of the joint posterior density estimate of $\theta=(\pi,\lambda$), its posterior mean (blue dot), posterior mode (green dot) and the true parameter value (red dot). The red curve is given by equation (\ref{eq.lambda}).}
        \label{fig:subfig6}
    \end{subfigure}

    \vspace{1cm}

    \begin{subfigure}[b]{0.45\textwidth}
        \includegraphics[width=\textwidth]
{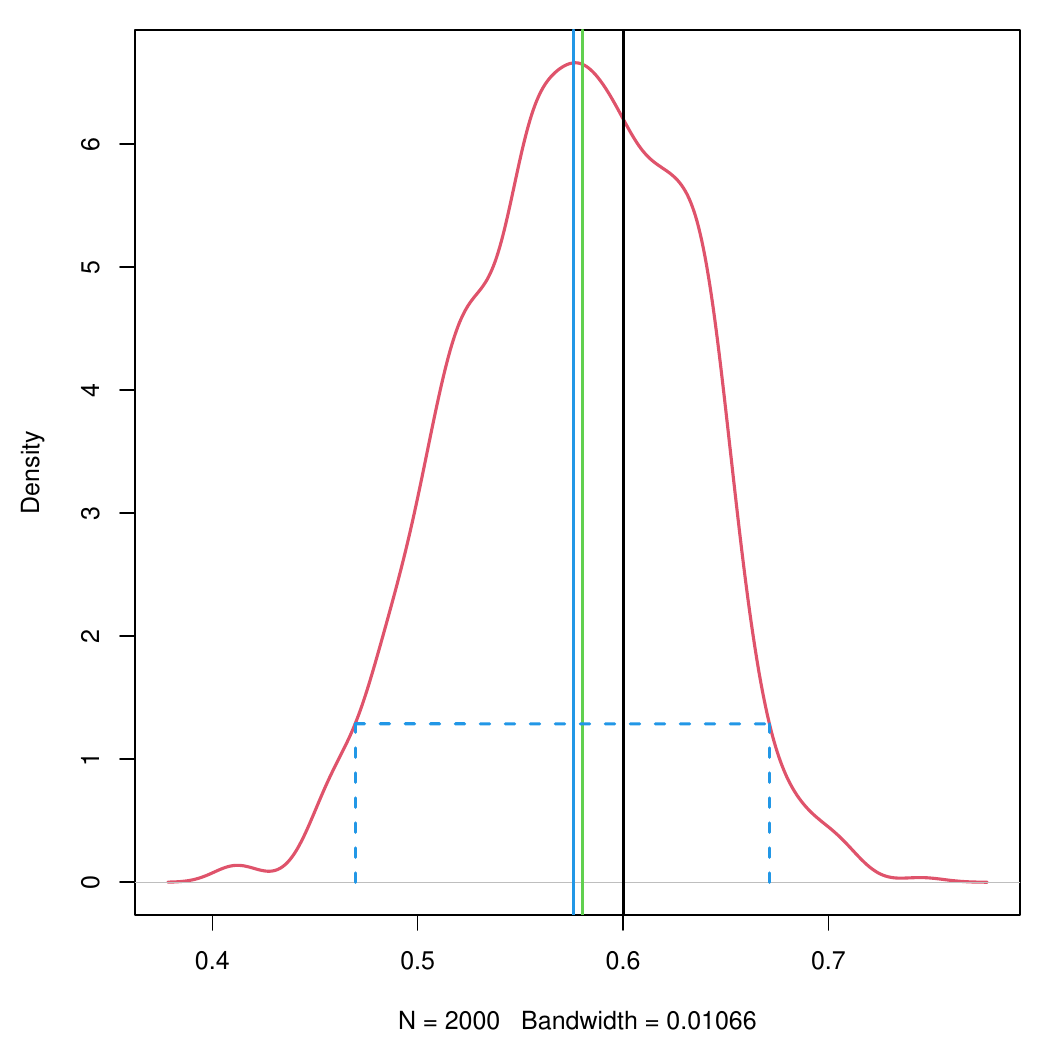}
        \caption{Marginal posterior density estimate of $\lambda$ with its posterior mean (blue line), posterior mode (green line), the true parameter value (black line) and the 95\% HPD credible set.}
        \label{fig:subfig7}
    \end{subfigure}
    \hfill
    \begin{subfigure}[b]{0.45\textwidth}
        \includegraphics[width=\textwidth]
{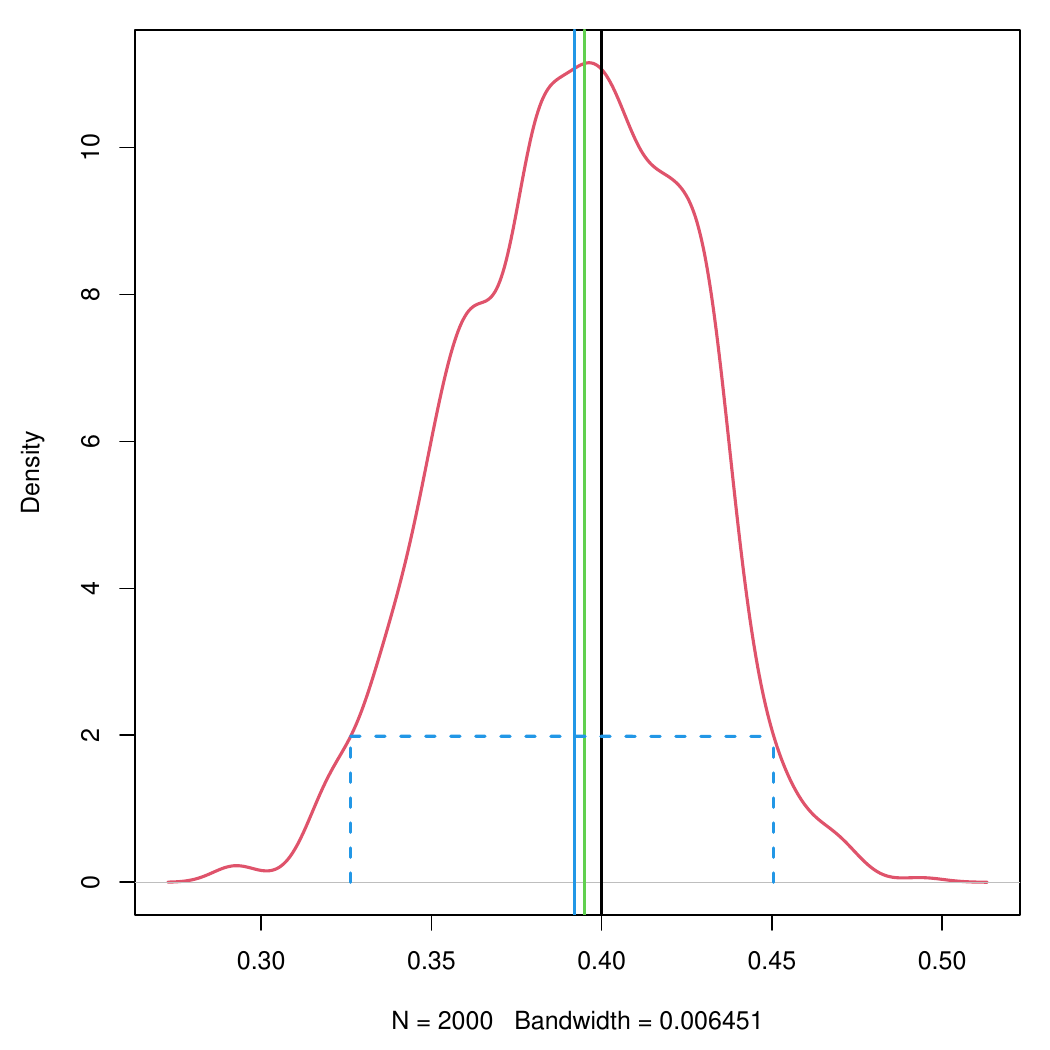}
        \caption{Marginal posterior density estimate of $\pi$ with its posterior mean (blue line), posterior mode (green line), the true parameter value (black line) and the 95\% HPD credible set.}
        \label{fig:subfig8}
    \end{subfigure}

    \caption{Case: $\pi=0.4$ and $\lambda=0.6$.}
    \label{fig:main2}
\end{figure}

\begin{figure}[ht]
    \centering
    % Primera fila de imágenes
    \begin{subfigure}[b]{0.45\textwidth}
        \includegraphics[width=\textwidth]
{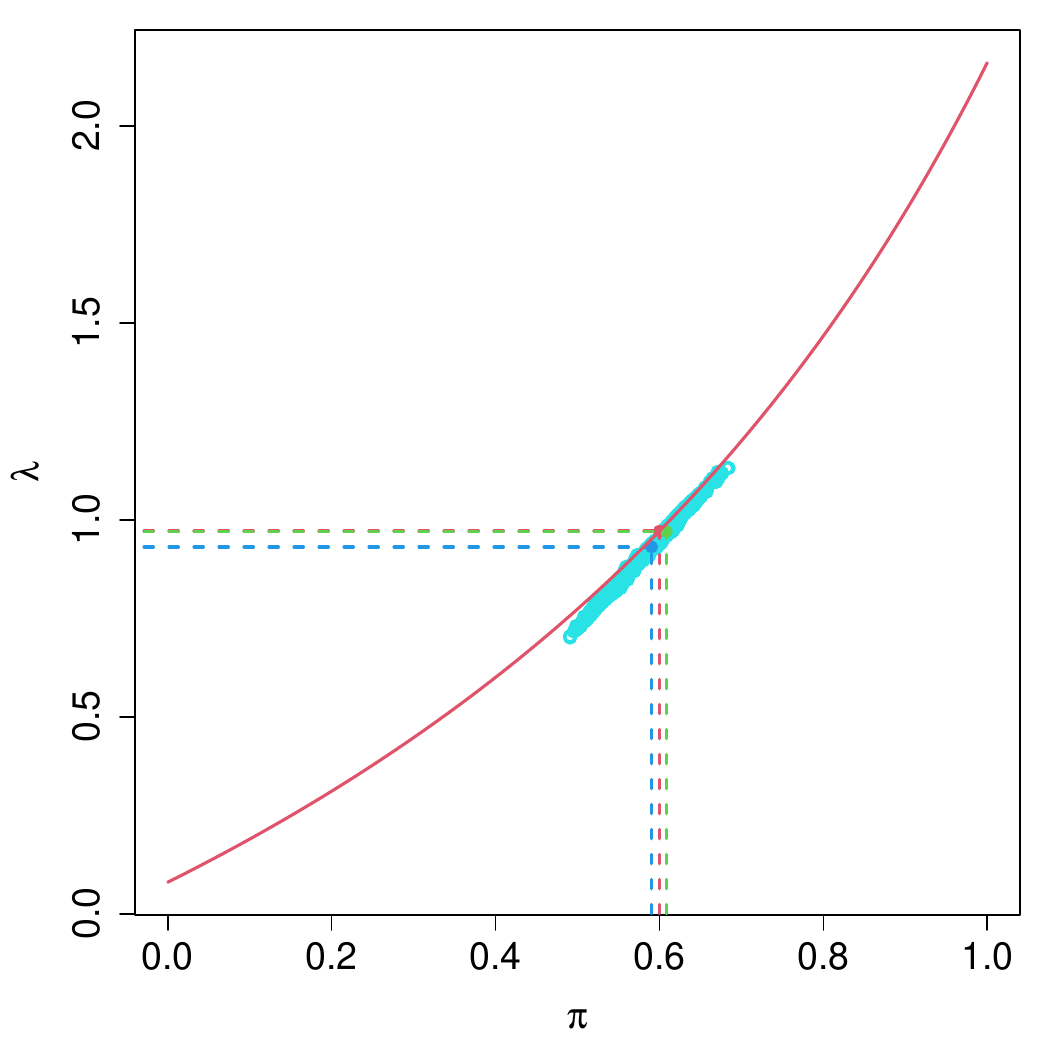}
        \caption{Scatterplot of the particles sampled from the joint posterior density estimate of $\theta=(\pi,\lambda$), its posterior mean (blue dot), posterior mode (green dot) and the true parameter value (red dot). The red curve is given by equation (\ref{eq.lambda}).}
        \label{fig:subfig9}
    \end{subfigure}
    \hfill
    \begin{subfigure}[b]{0.45\textwidth}
        \includegraphics[width=\textwidth]
{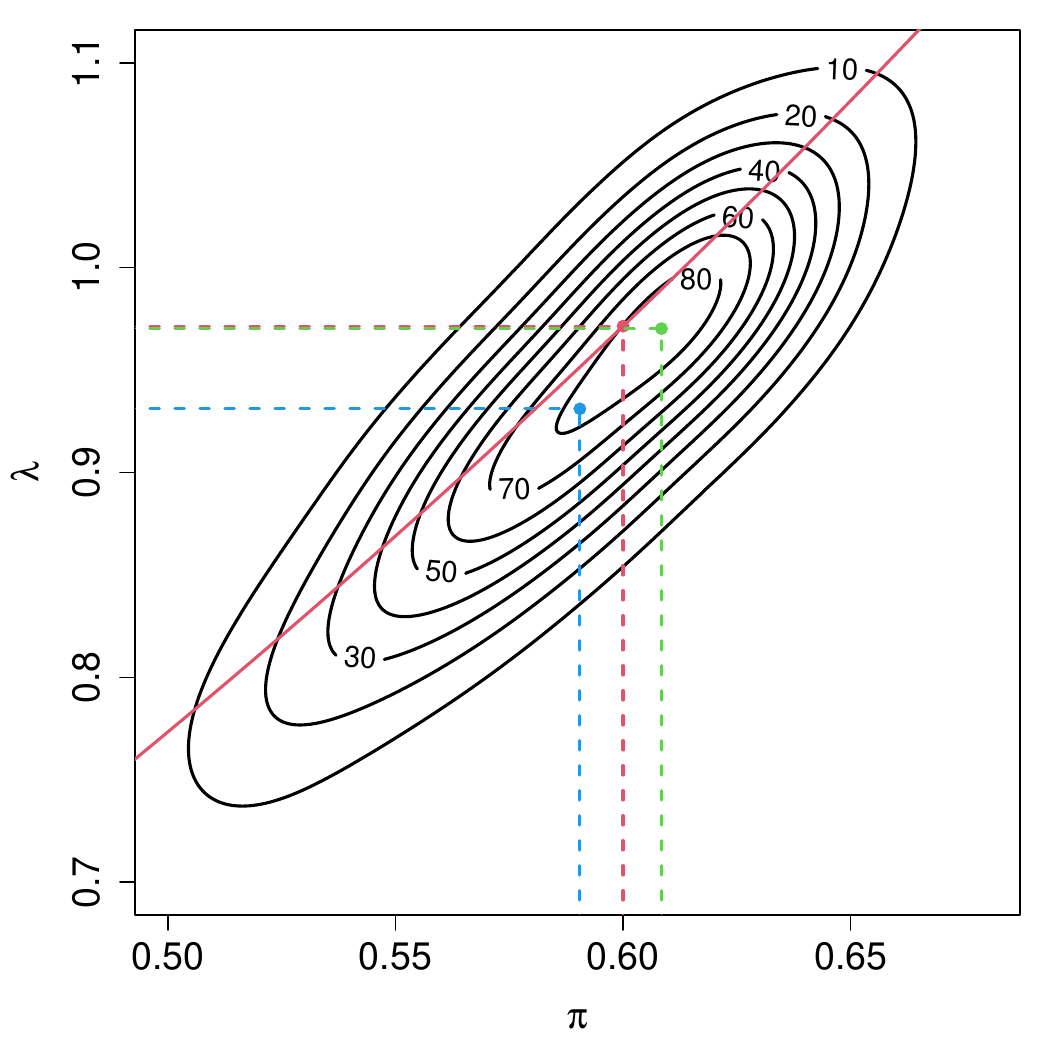}
        \caption{Contour plot of the joint posterior density estimate of $\theta=(\pi,\lambda$), its posterior mean (blue dot), posterior mode (green dot) and the true parameter value (red dot). The red curve is given by equation (\ref{eq.lambda}).}
        \label{fig:subfig10}
    \end{subfigure}

    \vspace{1cm}

    \begin{subfigure}[b]{0.45\textwidth}
        \includegraphics[width=\textwidth]
{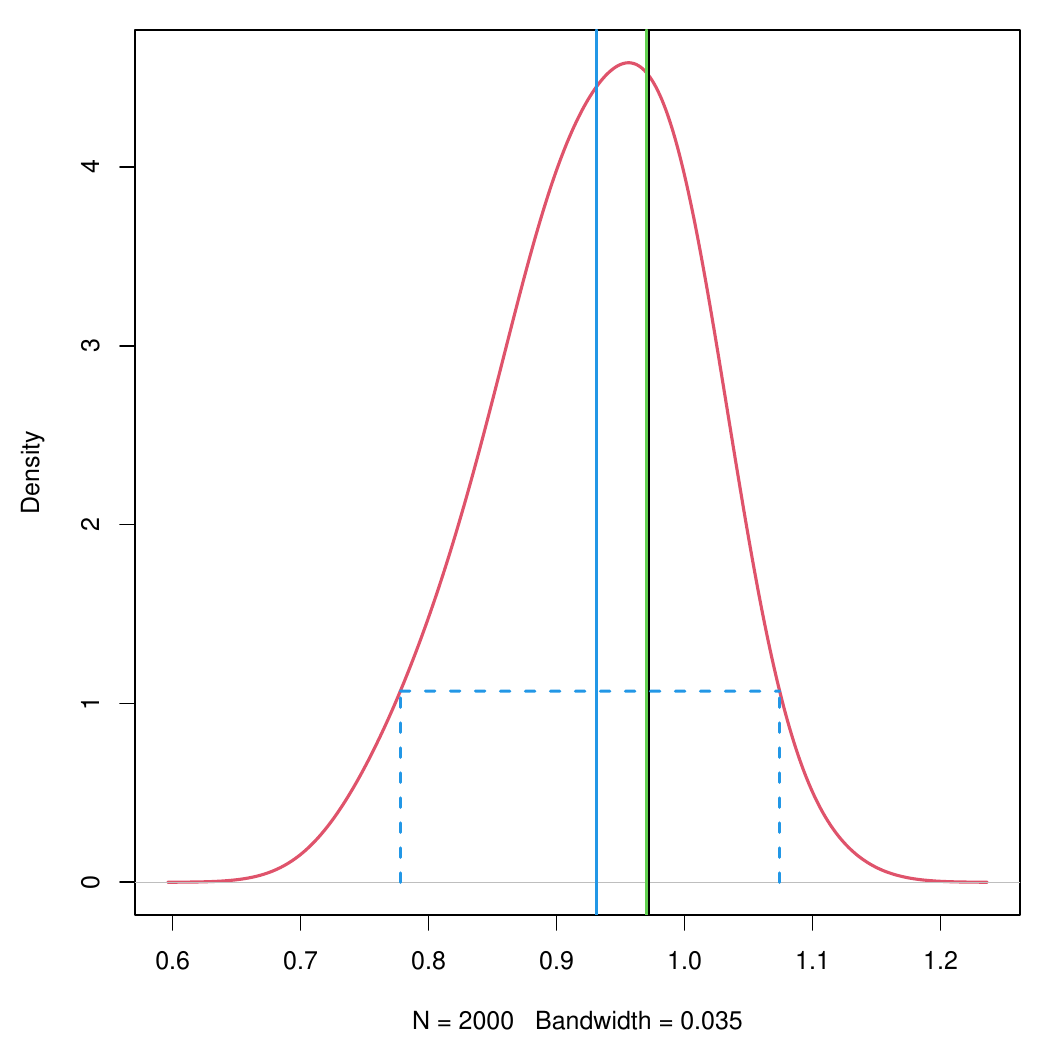}
        \caption{Marginal posterior density estimate of $\lambda$ with its posterior mean (blue line), posterior mode (green line), the true parameter value (black line) and the 95\% HPD credible set.}
        \label{fig:subfig11}
    \end{subfigure}
    \hfill
    \begin{subfigure}[b]{0.45\textwidth}
        \includegraphics[width=\textwidth]
{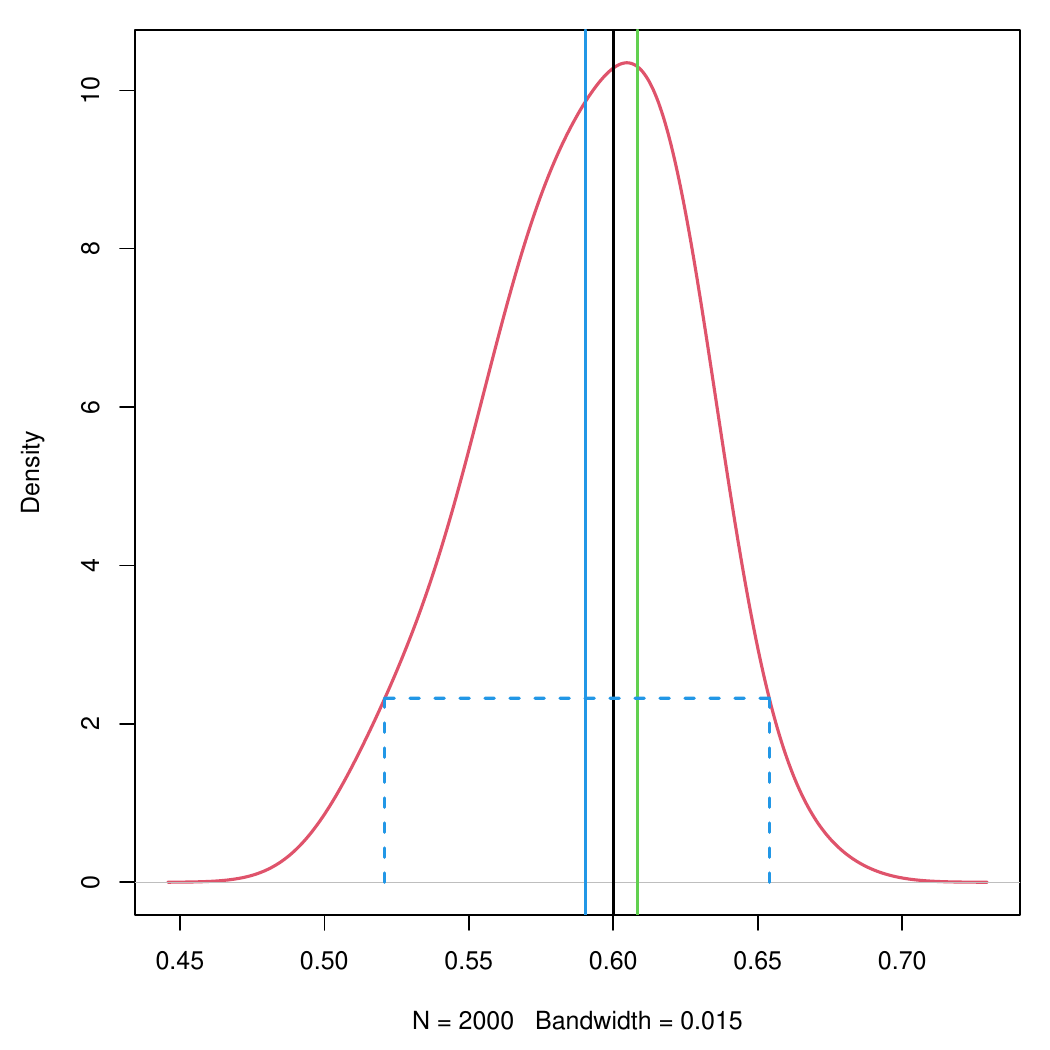}
        \caption{Marginal posterior density estimate of $\pi$ with its posterior mean (blue line), posterior mode (green line), the true parameter value (black line) and the 95\% HPD credible set.}
        \label{fig:subfig12}
    \end{subfigure}

    \caption{Case: $\pi=0.6$ and $\lambda=0.97$.}
    \label{fig:main3}
\end{figure}

\begin{figure}[ht]
    \centering
    % Primera fila de imágenes
    \begin{subfigure}[b]{0.45\textwidth}
        \includegraphics[width=\textwidth]
{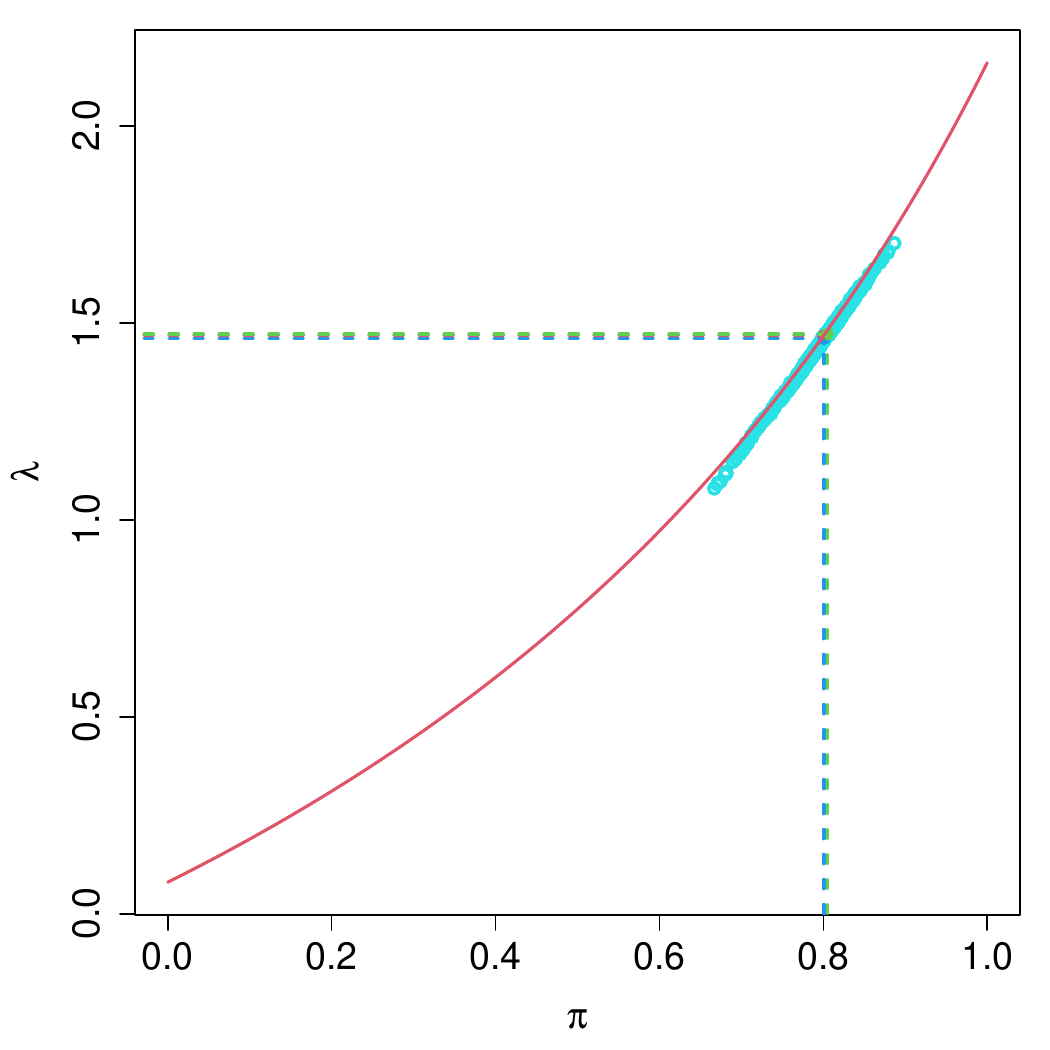}
        \caption{Scatterplot of the particles sampled from the joint posterior density estimate of $\theta=(\pi,\lambda$), its posterior mean (blue dot), posterior mode (green dot) and the true parameter value (red dot). The red curve is given by equation (\ref{eq.lambda}).}
        \label{fig:subfig13}
    \end{subfigure}
    \hfill
    \begin{subfigure}[b]{0.45\textwidth}
        \includegraphics[width=\textwidth]
{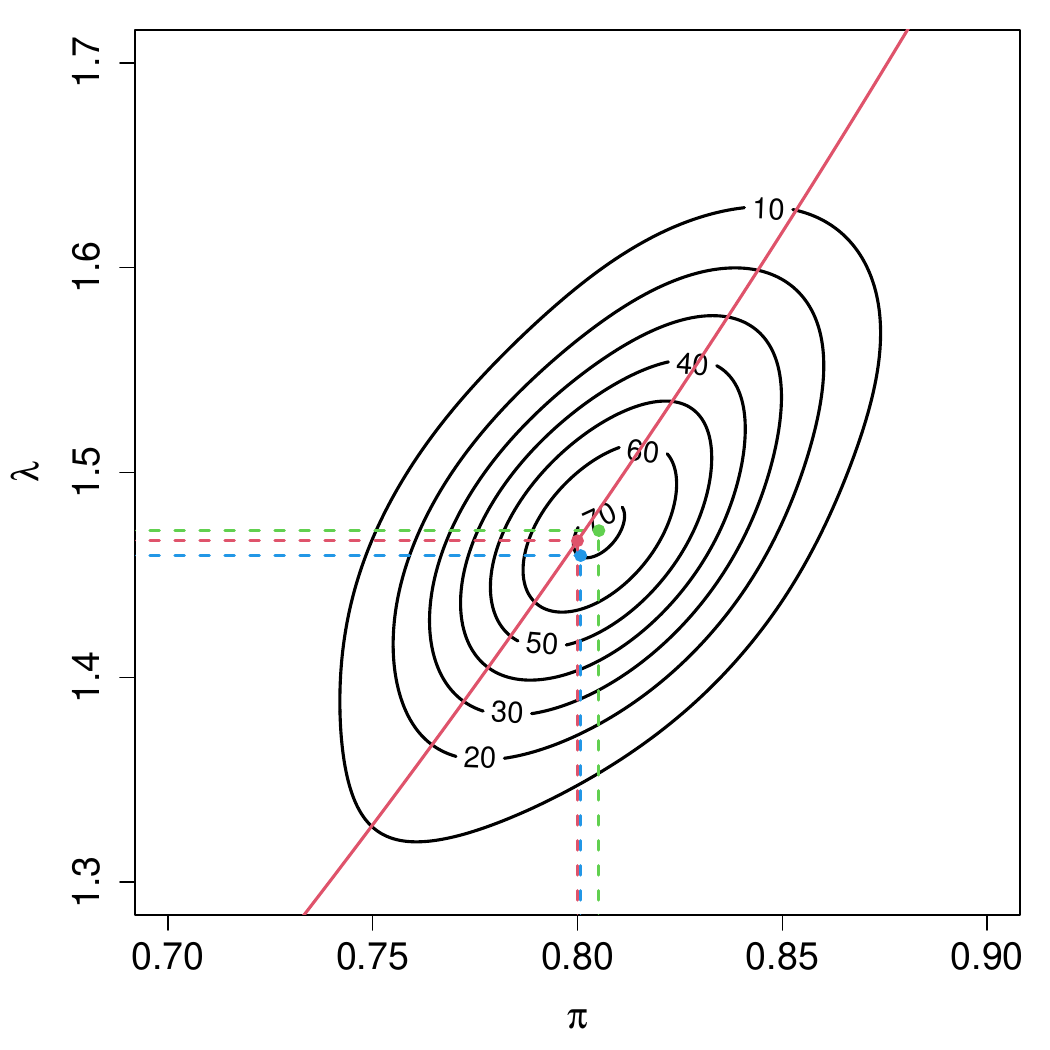}
        \caption{Contour plot of the joint posterior density estimate of $\theta=(\pi,\lambda$), its posterior mean (blue dot), posterior mode (green dot) and the true parameter value (red dot). The red curve is given by equation (\ref{eq.lambda}).}
        \label{fig:subfig14}
    \end{subfigure}

    \vspace{1cm}

    \begin{subfigure}[b]{0.45\textwidth}
        \includegraphics[width=\textwidth]
{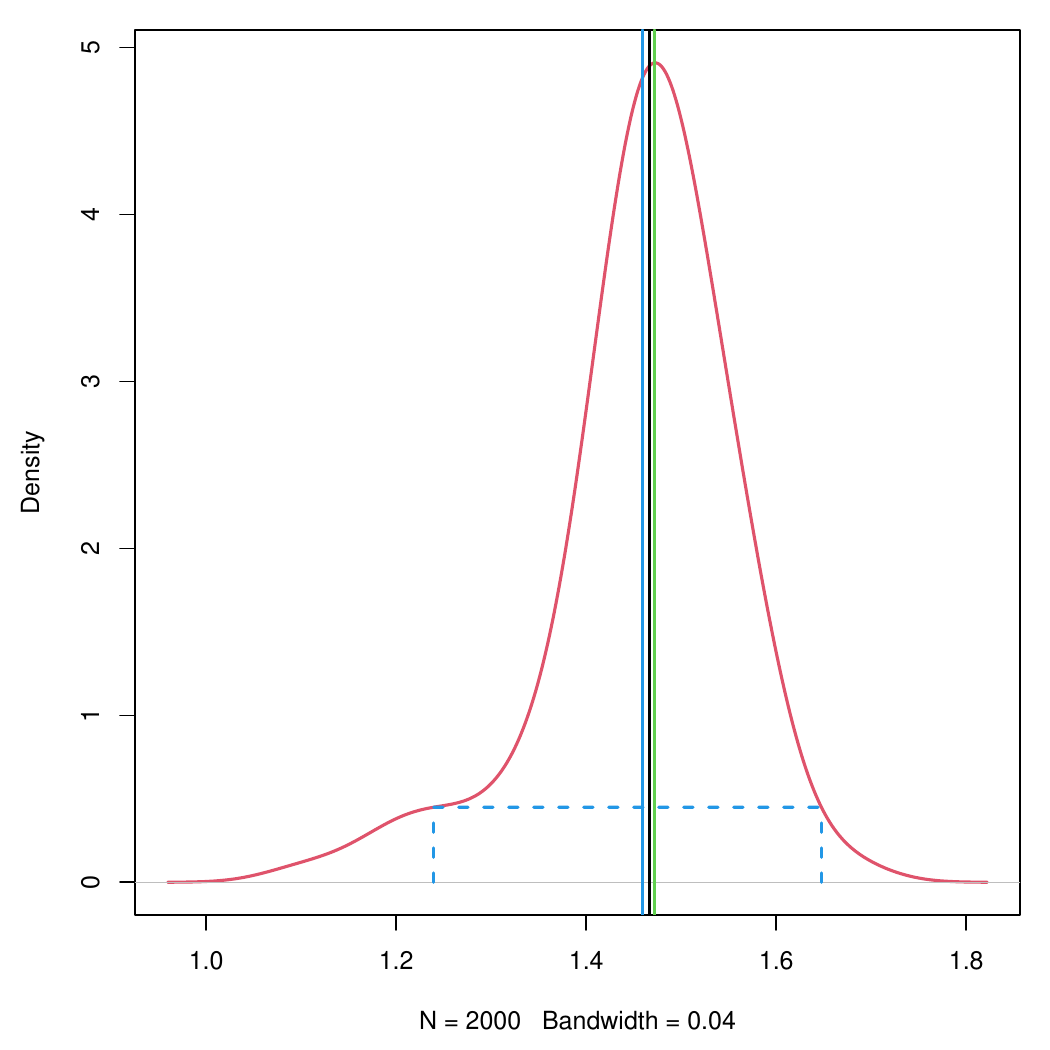}
        \caption{Marginal posterior density estimate of $\lambda$ with its posterior mean (blue line), posterior mode (green line), the true parameter value (black line) and the 95\% HPD credible set.}
        \label{fig:subfig15}
    \end{subfigure}
    \hfill
    \begin{subfigure}[b]{0.45\textwidth}
        \includegraphics[width=\textwidth]
{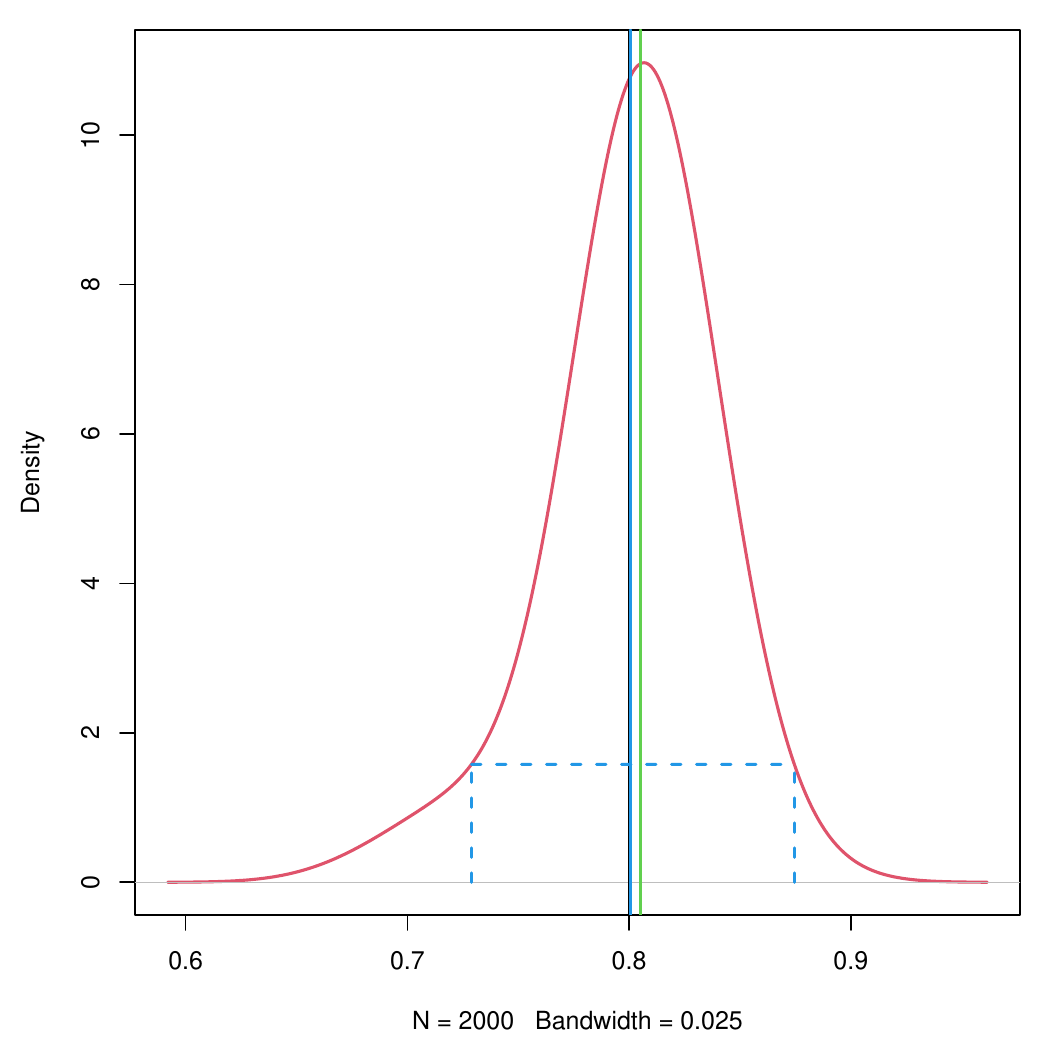}
        \caption{Marginal posterior density estimate of $\pi$ with its posterior mean (blue line), posterior mode (green line), the true parameter value (black line) and the 95\% HPD credible set.}
        \label{fig:subfig16}
    \end{subfigure}

    \caption{Case: $\pi=0.8$ and $\lambda=1.47$.}
    \label{fig:main4}
\end{figure}

In Figures \ref{fig:main1} - \ref{fig:main4} we show the joint posterior distribution of $\theta$, along with the marginal posterior distributions of $\pi$ and $\lambda$, including high posterior density (HPD) credible sets. In all cases, we obtain accurate estimates of both parameters. These estimates are evaluated under the squared error loss function — using the posterior means — and under the 0–1 loss function — using the posterior modes. Unlike the frequentist estimates for 
$\pi$ proposed in \cite{Meester2002}, our estimates appear to perform quite well even when only a small number of generations are observed -- 30 in our simulated study.

\begin{figure}[ht]
    \centering
    % Primera fila de imágenes
    \begin{subfigure}[b]{0.45\textwidth}
        \includegraphics[width=\textwidth]
{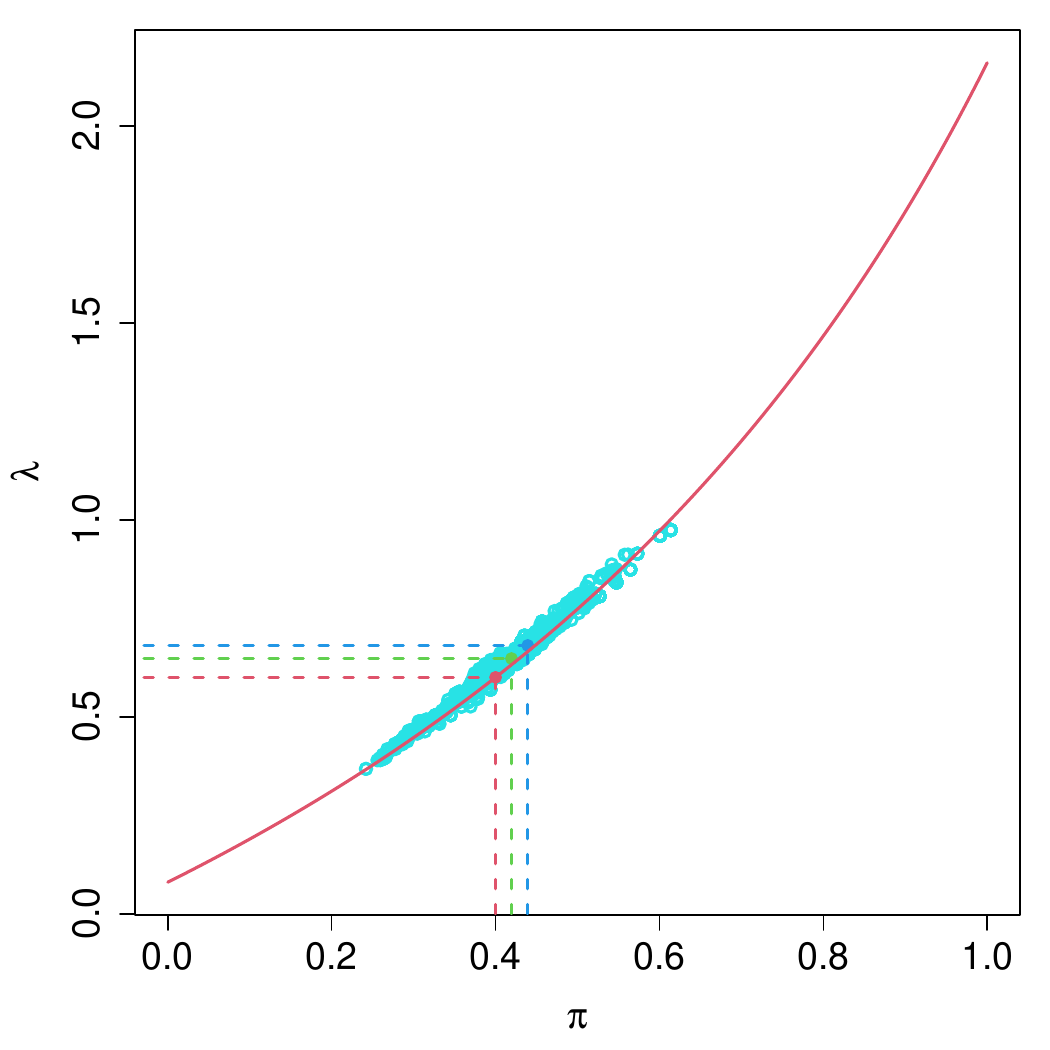}
        \caption{Scatterplot of the particles sampled from the joint posterior density estimate of $\theta=(\pi,\lambda$), its posterior mean (blue dot), posterior mode (green dot) and the true parameter value (red dot). The red curve is given by equation (\ref{eq.lambda}).}
        \label{fig:subfig17}
    \end{subfigure}
    \hfill
    \begin{subfigure}[b]{0.45\textwidth}
        \includegraphics[width=\textwidth]
{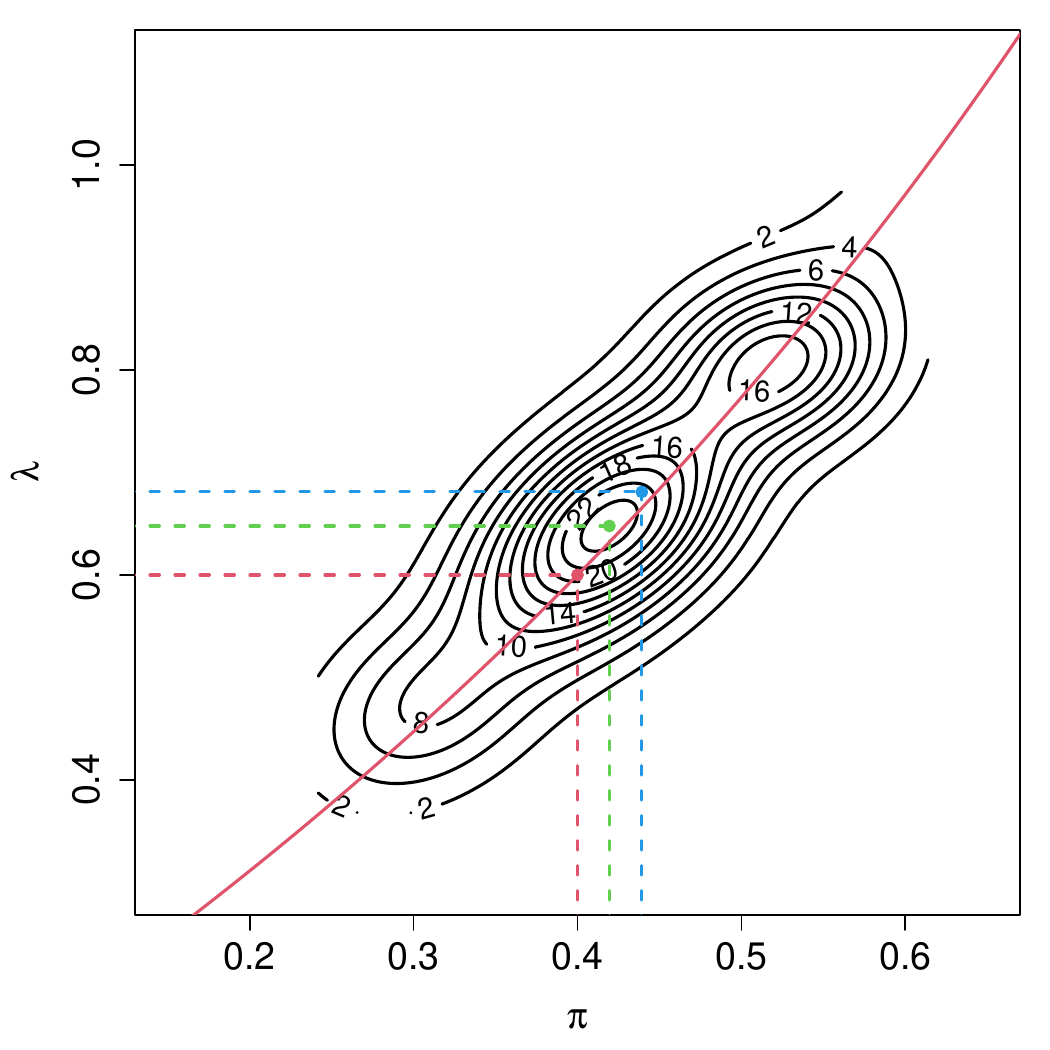}
        \caption{Contour plot of the joint posterior density estimate of $\theta=(\pi,\lambda$), its posterior mean (blue dot), posterior mode (green dot) and the true parameter value (red dot). The red curve is given by equation (\ref{eq.lambda}).}
        \label{fig:subfig18}
    \end{subfigure}

    \vspace{1cm}

    \begin{subfigure}[b]{0.45\textwidth}
        \includegraphics[width=\textwidth]
{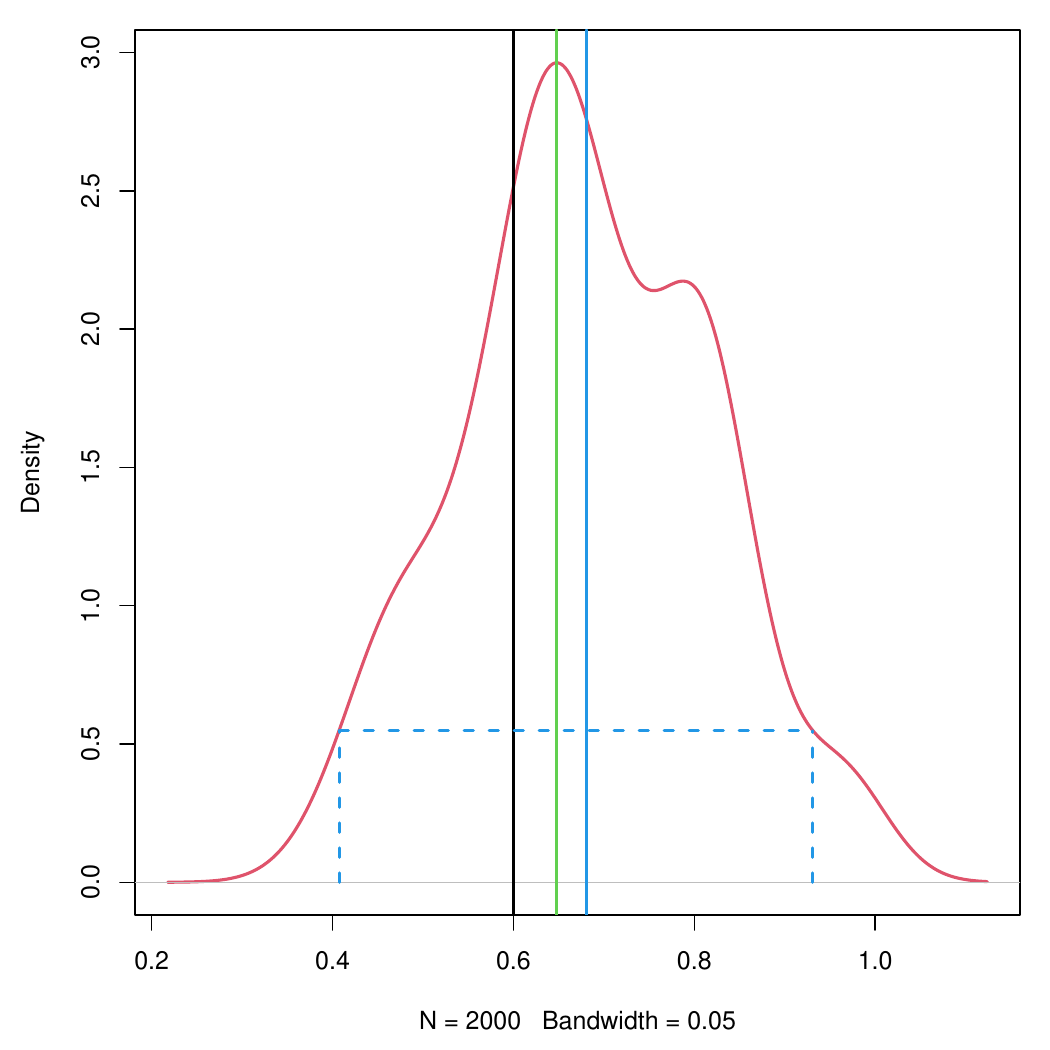}
        \caption{Marginal posterior density estimate of $\lambda$ with its posterior mean (blue line), posterior mode (green line), the true parameter value (black line) and the 95\% HPD credible set.}
        \label{fig:subfig19}
    \end{subfigure}
    \hfill
    \begin{subfigure}[b]{0.45\textwidth}
        \includegraphics[width=\textwidth]
{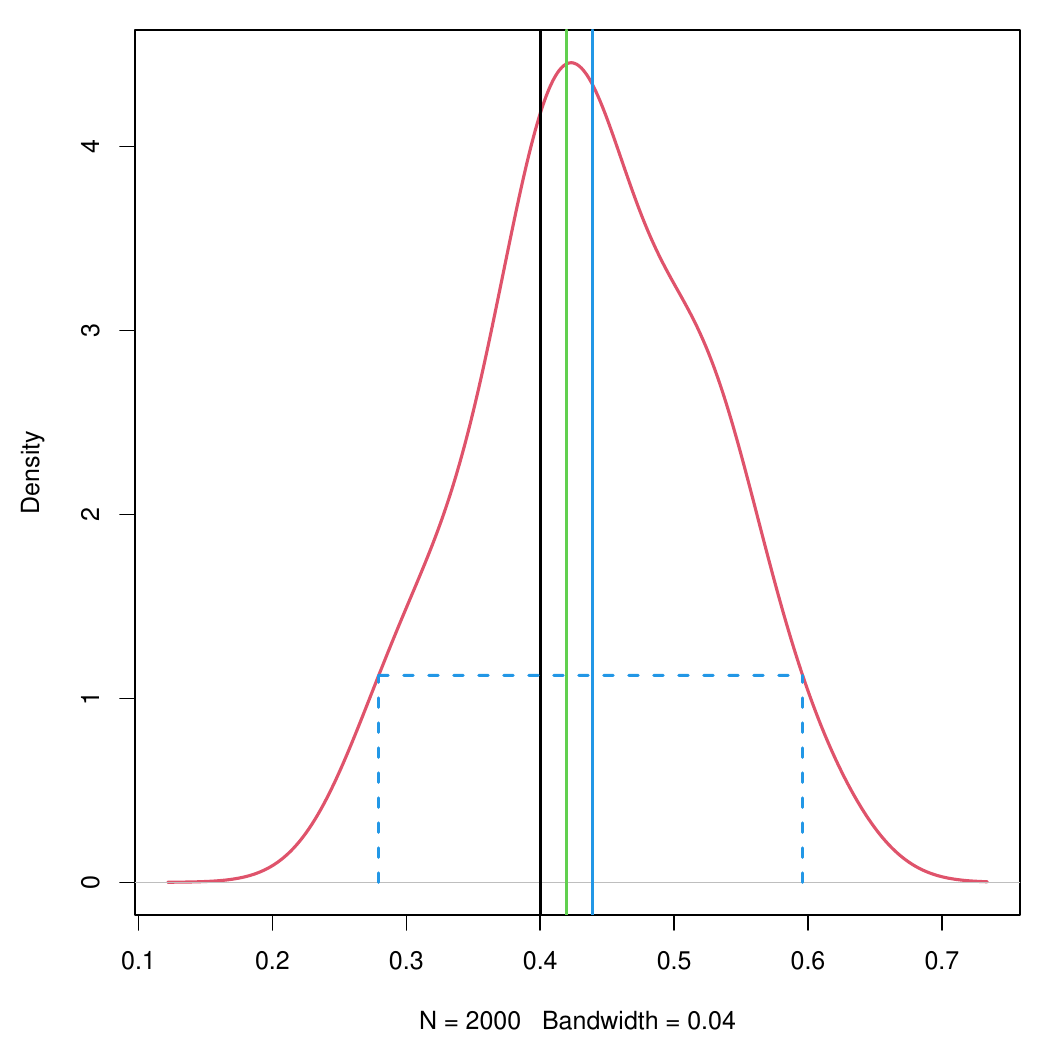}
        \caption{Marginal posterior density estimate of $\pi$ with its posterior mean (blue line), posterior mode (green line), the true parameter value (black line) and the 95\% HPD credible set.}
        \label{fig:subfig20}
    \end{subfigure}

    \caption{Case: $\pi=0.4$ and $\lambda=0.6$ with $x_0\sim U\{50,\ldots,200\}$.}
    \label{fig:main5}
\end{figure}

We also analyse the case $\theta = (\pi, \lambda) = (0.4, 0.6)$, assuming that the initial number of infected individuals, $x_0$, is an additional unknown parameter in the model. In this scenario, a discrete uniform prior over the set $\{50, \ldots, 200\}$ was assigned to $x_0$. The corresponding results are presented in Figure~\ref{fig:main5}. Although an increase in the dispersion of the posterior distributions is observed, the parameter estimates remain sufficiently accurate.

Finally, it is important to highlight that, from a computational perspective, the method based on the Liu–West algorithm outperforms alternative approaches such as Markov Chain Monte Carlo (MCMC) and Approximate Baye\-sian Computation (ABC), both in terms of computational time and memory efficiency, as it requires less data to be stored.

All the examples have been developed making use of the programming language and statistical software  \textbf{\textsf{R}}. 

\section*{Acknowledgements}
M. Gonz\'alez and I. del Puerto are supported by grant PID2023--152359NB--I00 funded by MICIU/AEI/10.13039/501100011033 and,  by
“ERDF/EU”.


\end{document}